\documentclass[usenatbib,usegraphicx,useAMS]{mn2e}
\pdfoutput=1
\usepackage{times} 
\usepackage{amsmath}
\usepackage{amssymb}
\usepackage{booktabs}
\usepackage[usenames,dvipsnames]{color}
\usepackage{dsfont}
\usepackage{graphicx}
\usepackage{url}
\usepackage{listings} 
\usepackage[point]{rccol}
\usepackage{soul}
\usepackage{subfig}
\usepackage{verbatim}
\usepackage{natbib}
\usepackage{hyperref}

\allowdisplaybreaks[1]

\newcommand{\App}[1]{Appendix~\ref{#1}}

\newcommand{\Eq}[1]{Eq.~(\ref{#1})}

\newcommand{\Fig}[1]{Fig.~\ref{#1}}
\newcommand{\Sec}[1]{\S\ref{#1}}
\newcommand{\Tab}[1]{Table~\ref{#1}}

\def\adaptahop    {{\sc AdaptaHOP}}
\def\bc    {{\sc BC03}}
\def\bpass    {{\sc Bpass}}
\def\cloudy    {{\sc Cloudy}}
\def\galaxev    {{\sc Galaxev}}
\def\music    {{\sc Music}}
\def\ramses    {{\sc Ramses}}
\def\ramsesrt  {{\sc Ramses-RT}}
\def\sphinx  {{\sc Sphinx}}

\def\stromgren {Str\"omgren}

\def\Sphsmhr {{\sc S05\_512}}
\def\Sphsmhrbp {{\sc S05\_512\_Binary}}
\def\Sphsm {{\sc S05\_256}}
\def\Sphlg {{\sc S10\_512}}
\def\Sphlgbp {{\sc S10\_512\_Binary}}

\def\hi    {{\rm{H\textsc{i}}}}
\def\hii   {\rm{H\textsc{ii}}}
\def\hei    {{\rm{He\textsc{i}}}}
\def\heii    {{\rm{He\textsc{ii}}}}
\def\nh {n_{\rm{H}}}
\def\Nh {N_{\rm{H}}}
\def\Nhi {N_{\rm{H\textsc{i}}}}

\def\xhii {x_{\rm{H \scriptscriptstyle II}}}

\def\xheii {x_{\rm{He \scriptscriptstyle II}}}
\def\xheiii {x_{\rm{He \scriptscriptstyle III}}}
\def\cci {{\rm{cm}}^{-3}}   
\def\ccitwod {{\rm{cm}}^{-2}}   
\def\cs {\rm{cm}^{2}}       
\def\Msun {{\rm{M}_{\odot}}}
\def\sm {\rm{s}^{-1}}       

\def\cred {\tilde c}      
\def\csavg {\sigma}           
\def\csound {c_{\rm{s}}}
\def\Dpmax {\Delta p_{\rm max}}

\def\Dx {\Delta x}

\def\Dxmax {\Delta x_{\rm max}}
\def\Dxmin {\Delta x_{\rm min}}
\def\eavg {\bar{\epsilon}}   

\def\eSN {E_{\rm SN}}
\def\eOne {E_{51}}
\def\etaSN {\eta_{\rm SN}}

\def\fc {f_c}
\def\fe {f_{\rm e}}
\def\fesc {f_{\rm esc}}
\def\feschundred {f_{\rm esc,100}}
\def\fescsr {f'_{\rm esc}}

\def\fnbor {f_{\rm nbor}}

\def\Gammahi {\Gamma_{\rm HI}}
\def\Hubble {h}
\def\IonLum {{\mathcal{L}}}
\def\IonLumMass {\mathcal{L}_M}
\def\IonLumVol {\tilde{\mathcal{L}}}
\def\kb {k_{\rm B}}
\def\Lbox {L_{\rm box}}
\def\levelmax {\ell_{\rm max}}
\def\levelmin {\ell_{\rm min}}
\def\LJeans {\lambda_{\rm J}}
\def\LJeansT {\lambda_{\rm J, turb}}
\def\LumFifteen {{L_{1500}}}
\def\Lumint {\Pi}
\def\Mab {M_{\rm AB}}
\def\mDM {m_{\rm DM}}

\def\MDMcell {M_{\rm DM, cell}}

\def\Mbaryonscell {M_{\rm baryons, cell}}

\def\Mvir {M_{\rm vir}}

\def\mSN {m_{\rm SN}}

\def\mW {m_{\rm W}}
\def\mstar {m_{*}}
\def\Mstar {M_{*}}
\def\Mtwoh {M_{\rm 200}}
\def\Mvir {M_{\rm vir}}

\def\mp {m_{\rm p}}
\def\Nphot {\mathcal{N}}

\def\Omegab {\Omega_{\rm b}}
\def\Omegam {\Omega_{\rm m}}

\def\Qhi {Q_{\rm HI}}

\def\rS {r_{\rm{S}}}        

\def\Rvir {R_{\rm vir}}
\def\sfeff {\epsilon_{*}}

\def\siggas {\sigma_{\rm gas}}

\def\Sigstar {\Sigma_{*}}

\def\tff {t_{\rm{ff}}}

\def\Zsun {Z_{\odot}}           

\mathchardef\mhyphen="2D

\long\def\symbolfootnote[#1]#2{\begingroup%
\def\thefootnote{\fnsymbol{footnote}}\footnote[#1]{#2}\endgroup}

\begin{document}

\title[The \sphinx{} simulations of reionization]{The SPHINX Cosmological
  Simulations of the First Billion Years: \\the Impact of Binary Stars
  on Reionization\thanks{\url{https://sphinx.univ-lyon1.fr/}}}

\author[Rosdahl et al.] {Joakim Rosdahl$^{1}$\thanks{E-mail:
    karl-joakim.rosdahl@univ-lyon1.fr},
  Harley Katz$^{2,3}$,
  J\'er\'emy Blaizot$^{1}$,
  Taysun Kimm$^{4,3}$,
  \newauthor L\'eo Michel-Dansac$^{1}$,
  Thibault Garel$^{1}$,
  Martin Haehnelt$^{3}$,
  \newauthor Pierre Ocvirk$^{5}$,
  and Romain Teyssier$^{6}$. \\
  $^1$Univ Lyon, Univ Lyon1, Ens de Lyon, CNRS, Centre de Recherche
  Astrophysique de Lyon UMR5574, F-69230, Saint-Genis-Laval, France \\
  $^2$Sub-department of Astrophysics, University of Oxford,
  Keble Road, Oxford OX1 3RH, UK \\
  $^3$Kavli Institute for Cosmology and Institute of Astronomy,
  Madingley Road, Cambridge CB3 0HA, UK \\
  $^4$Department of Astronomy, Yonsei University, 50 Yonsei-ro,
  Seodaemun-gu, Seoul 03722, Republic of Korea \\
  $^5$Observatoire Astronomique de Strasbourg, Universit\'e de
  Strasbourg, CNRS UMR 7550, 11 rue de l'Universit\'e, 67000
  Strasbourg, France \\
  $^6$Institute for Computational Science, University of Z\"urich,
  Winterthurerstrasse 190, CH-8057 Z\"urich, Switzerland}

\maketitle
\begin{abstract}
  We present the \sphinx{} suite of cosmological adaptive mesh
  refinement simulations, the first radiation-hydrodynamical
  simulations to simultaneously capture large-scale reionization and
  the escape of ionizing radiation from thousands of resolved
  galaxies. Our $5$ and $10$ co-moving Mpc volumes resolve haloes down
  to the atomic cooling limit and model the inter-stellar medium with
  better than $\approx10$ pc resolution. The project has numerous
  goals in improving our understanding of reionization and making
  predictions for future observations. In this first paper we study
  how the inclusion of binary stars in computing stellar luminosities
  impacts reionization, compared to a model that includes only single
  stars. Owing to the suppression of galaxy growth via strong
  feedback, our galaxies are in good agreement with observational
  estimates of the galaxy luminosity function. We find that binaries
  have a significant impact on the timing of reionization: with
  binaries, our boxes are $99.9$ percent ionized by volume at
  $z\approx 7$, while without them our volumes fail to reionize by
  $z=6$. These results are robust to changes in volume size,
  resolution, and feedback efficiency. The escape of ionizing
  radiation from individual galaxies varies strongly and
  frequently. On average, binaries lead to escape fractions of
  $\approx 7-10$ percent, about $3$ times higher than with single
  stars only.  The higher escape fraction is a result of a shallower
  decline in ionizing luminosity with age, and is the primary reason
  for earlier reionization, although the higher integrated luminosity
  with binaries also plays a sub-dominant role.
\end{abstract}

\section{Introduction} \label{Intro.sec} 

The formation of the first galaxies marks the end of the dark ages and
the beginning of the Epoch of Reionization (EoR). Radiation from the
first generations of stars, hosted by the first galaxies, heated the
surrounding inter-galactic gas via photo-ionization. As the ionized
hydrogen ($\hii$) bubbles grew and percolated, the whole Universe was
transformed from a dark, cold, neutral state into a hot ionized one:
reionization was completed. This last major transition of the Universe
is at the limit of our observational capabilities and is a key science
driver of the foremost upcoming telescopes, such as the James Webb
Space Telescope (JWST) and the Square Kilometre Array (SKA).

Cosmological simulations are an indispensable tool to disentangle the
complex and non-linear interplay of physical mechanisms leading to
reionization, a `loop' encompassing an enormous range of physical
scales: of gravitational collapse of dark matter into haloes, the
condensation of gas via radiative cooling into galaxies at the centers
of those haloes, it’s eventual collapse into stars, which is slowed
down by feedback, the emission of ionizing radiation from those stars,
the propagation of the radiation through the inter-stellar medium
(ISM), the circum-galactic medium (CGM) and the inter-galactic medium
(IGM), and its ability to suppress gas cooling and thus star
formation.

Two major challenges face those seeking to understand reionization
with simulations, compared with simulations of galaxy evolution at
low-redshift. First is the need for radiation-hydrodynamics (RHD) to
explicitly model the interplay of ionizing radiation and
gas. Traditional cosmological simulations use pure
gravito-hydrodynamics, typically applying homogeneous ultra-violet
(UV) background radiation instead of the more computationally
expensive hydrodynamically coupled radiative transfer. However, to
model the reionization process self-consistently, the radiation should
not be ignored or applied in post-processing. The second challenge is
capturing the range of scales involved. Reionization is a large-scale
process that should preferentially be modelled on
cosmological-homogeneity scales, or volume widths of $\gtrsim200$
co-moving Mpc (cMpc), in order to predict the patchiness of
reionization and a realistic average volume filling factor of ionized
gas that is unaffected by cosmic variance \citep{Iliev2013}. The
sources of reionization, however, form and emit radiation on
relatively tiny, sub-pc scales, and even if sub-pc scales are not
represented, predicting the transfer and escape of ionizing radiation
through the ISM requires at least a few pc resolution
\citep[e.g.][]{Kimm2014, Xu2016}.

There has been a surge of cosmological RHD reionization simulation
projects that have captured the large-scale reionization process
\citep{Gnedin2006, Finlator2011, Iliev2013, Gnedin2014, So2013a,
  Pawlik2016, Ocvirk2015, Chen2017}, with volume widths ranging from
$\sim10-600$ cMpc. However, the range of scales required to adequately
model \emph{both} the large-scale process and the production and
transfer of radiation through the ISM has been out of reach. Galaxies
are still largely unresolved in reionization simulations, with
physical resolution ranging from $\sim100$ pc \citep{Gnedin2014,
  Pawlik2016} for smaller volumes to $100$ kpc \citep{Iliev2013} for
the largest ones. Therefore, the escape fraction of ionizing radiation
from galaxies, $\fesc$, is fully or partly a free and adjustable
parameter. If galaxies are not resolved at all it sets the total
fraction of radiation getting out of galaxies. If galaxies are to some
degree resolved, the free parameter is a sub-resolution escape
fraction, $\fescsr$, setting the luminosity of stellar sources, and
the real escape fraction becomes a multiple of $\fescsr$, which sets
the production rate of ionizing photons, and the fraction of emitted
radiation that propagates out of galaxies \citep[e.g.][]{Pawlik2016}.

Cosmological \emph{zoom} simulations can achieve a much higher maximum
resolution than full cosmological simulations by focusing only on the
environment of one or a few haloes with a nested refinement
structure. In this type of simulation, the production and escape of
ionizing radiation in galaxies can be predicted in detail, even with
sub-parsec resolution. Furthermore, fewer assumptions regarding star
formation, feedback, and radiation-gas interactions are required.
Radiation-hydrodynamical zoom simulations have provided important
insight into reionization and particularly how $\fesc$ appears to
highly fluctuate and be strongly regulated by feedback
\citep{Wise2009, Kimm2014, Wise2014, Trebitsch2017, Xu2016, Kimm2017}.
However, the sacrifice is that the large-scale reionization process,
statistical averages, and the scatter (in e.g. $\fesc$ and ionizing
luminosities against galaxy properties) provided by the resolved
evolution of many haloes is lost. The cosmological environment
surrounding these few targeted well-resolved haloes remains at a
relatively low resolution and only serves to provide a realistic
background gravitational field.

The recent state-of-the-art is to connect these two types of
small-scale and large-scale simulations by using the predicted escape
of photons from the high-resolution zooms as inputs for the unresolved
escape of photons from galaxies in full reionization simulations,
allowing predictions to be made for the contributions of different
galaxy masses to reionization \citep{Xu2016, Chen2017}. Yet, this
approach still lacks the full coupling between the highly fluctuating
escape of ionizing radiation from these unresolved galaxies and the
galaxy dynamics, which is partially regulated by large-scale processes
of mergers and accretion. Full non-zoom cosmological RHD simulations
of reionization that capture these large scale processes and
simultaneously predict the production and escape of ionizing radiation
remain the ideal goal.

With recent increase in computational power and algorithmic advances
for radiation-hydrodynamics in the \ramsesrt{} code, this goal is now
within reach. Here, we present the \sphinx{}\footnote{In the spirit of
  the mythical Sphinx, we prefer to keep the acronym an enigma.}
suite of simulations, a series of cosmological volumes $5-10$ cMpc in
width where haloes are well resolved down to or below the atomic
cooling threshold (halo mass of
$\Mvir \approx 3 \times 10^{7} \ \Msun$; \citealt{Wise2014}) and a
maximum physical resolution of $11$ pc is reached in the ISM of
galaxies at $z=6$ (and higher at higher redshifts).

These volumes are still well below the cosmological homogeneity scale,
but we apply an initial conditions (ICs) selection technique that
allows us to minimise cosmic variance effects, model an accurate
ionizing photon production rate, and achieve reionization histories
that are well-converged with simulation volume. The simulation volumes
include thousands of star-forming galaxies, allowing for a statistical
understanding of escape fractions and ionizing luminosities for haloes
over a mass range of $\sim10^7-10^{11} \ \Msun$.

Our goals with the \sphinx{} project are numerous, including
understanding the main sources of reionization (e.g. halo masses,
environments, stellar population ages), the statistical behaviour of
$\fesc$, and the back-reaction \citep{Gnedin2014a} of radiation on the
formation of dwarf galaxies. Furthermore, we aim to make predictions
for the observational signatures of EoR galaxies for the JWST and to
better constrain the metal enrichment process of the high-redshift
IGM. These goals will be addressed in forthcoming papers. This pilot
\sphinx{} paper is dedicated to describing the numerical methods and
setup of our simulations, to comparing our simulated galaxies to
observations of the high-redshift Universe, and to addressing the
contribution of binary stars to reionization.

Recently, \cite{Stanway2016} demonstrated how accounting for the
interactions of binary stars in spectral energy distribution (SED)
models increases the total flux of ionizing radiation from metal-poor
stellar populations by tens of percent, compared to models that do not
include binaries. Around the same time, \cite{Ma2016} demonstrated an
additional, and perhaps more important, effect of binaries, which is
an increase in the escape of ionizing radiation from galaxies. They
post-processed cosmological zoom simulations with ray-tracing,
comparing two sets of SED models, with only single stars, and with
added binary stars, and found a factor $4-10$ higher $\fesc$ with
binary stars included (with $\fesc\gtrsim20\%$ for binaries).

This ties to the regulation of $\fesc$ by stellar feedback.  In the
first few Myr after the birth of a stellar population, very little
ionizing radiation escapes, even if the population is very luminous,
since the radiation is absorbed locally by the dense ISM.  The first
SN explosions at $\approx 3$ Myr coincide with a steep drop in the
production rate of ionizing photons, because the most luminous stars
in the population are precisely those most massive ones that explode
first as SNe. So as the ISM starts to be cleared away by SN explosions
and the radiation begins to escape, the ionizing luminosity is rapidly
declining, resulting in an overall small fraction of the emitted
radiation escaping. Due to mass transfer and mergers between binary
companions, UV luminous stars exist at much later times in models that
include binaries compared to single stars only models. The luminosity
still declines after $3$ Myr, but at a slower rate, enhancing the
number of photons that can escape into the IGM \emph{after} the onset
of SN explosions.

\cite{Ma2016} predict a strong boost in $\fesc$ with the inclusion of
binary stars, but they apply the radiative transfer in post-processing
and only consider three galaxies from cosmological zoom simulations.
Therefore they do not consider the back reaction on the gas or predict
how binary stars affect the reionization history. In this paper, we
seek to verify and expand the results of \cite{Ma2016}, using directly
coupled radiation-hydrodynamics, and a full (non-zoom) cosmological
volume to provide the statistics of thousands of galaxies. This allows
us to directly probe whether and how the inclusion of binary stars
affects the evolution, observational properties, and escape fractions
of galaxies, as well as the timing and the process of reionization.

The setup of the paper is as follows. In section \ref{simulations.sec}
we present our simulation methods and setup (code, refinement,
radiative transfer, star formation, feedback, selection of initial
conditions). In section \ref{results.sec} we present our results,
first comparing our simulations with $z=6$ observables, then showing
the reionization histories resulting from binary and single star SED
models, and finally comparing overall escape fractions, with those
same models, from millions of stellar populations in thousands of
galaxies. We present our discussion of the greater impact of binaries
on reionization in section \ref{Discussion.sec} and conclude in
section \ref{Conclusions.sec} by highlighting upcoming work using the
\sphinx{} simulations.
 
\section{Simulations} \label{simulations.sec}

We use \ramsesrt{} \citep{Rosdahl2013, Rosdahl2015a}, which is an RHD
extension of the cosmological gravito-hydrodynamical code \ramses{}
\citep{Teyssier2002}\footnote{The public code, including all the RHD
  extensions used here, can be downloaded at
  \url{https://bitbucket.org/rteyssie/ramses}}, to solve the
interactions of dark matter, stellar populations, ionizing radiation,
and baryonic gas via gravity, hydrodynamics, radiative transfer, and
non-equilibrium radiative cooling/heating, on a three-dimensional
adaptive mesh. For the hydrodynamics, we use the HLLC Riemann solver
\citep{Toro1994} and the MinMod slope limiter to construct gas
variables at cell interfaces from their cell-centred values. To close
the relation between gas pressure and internal energy, we use an
adiabatic index $\gamma=5/3$, which is appropriate for an ideal
monatomic gas. The dynamics of collisionless DM and stellar particles
are evolved with a particle-mesh solver and cloud-in-cell
interpolation \citep{Guillet2011}. The advection of radiation between
cells is solved with a first-order moment method, using the fully
local M1 closure for the Eddington tensor \citep{Levermore1984} and
the Global-Lax-Friedrich flux function for constructing the inter-cell
radiation field.

In the following sub-sections, we describe the set-up of our
simulations (initial conditions, adaptive refinement, ionizing
radiation, thermochemistry), our main sub-grid model components (star
formation and supernova feedback), and the halo finder utilised in the
analysis. For reference, the main simulation parameters are listed in
\Tab{sim_params.tbl}.

\begin{table}
  \begin{center}
  \caption
  {Simulation parameters, by name, value, and description.}
  \label{sim_params.tbl}
  \begin{tabular}{r|l|r}
    \toprule
    Name & Value & Comments \\
    \midrule
    $\Omega_{\rm m}$ & $0.3175$ & Initial conditions (ICs) mass density  \\
    $\Omega_{\Lambda}$ & $0.6825$ & ICs dark energy density  \\
    $\Omega_{\rm b}$ & $0.049$ & ICs baryon density  \\
    $\sigma_8$ & $0.83$ & ICs amplitude of galaxy fluctuations  \\
    $H_0$ & $67.11 \ \ {\rm km/s/Mpc}$ & Hubble constant   \\
    \midrule
    $X$ & $0.74$ 
                 & Hydrogen mass fraction \\
    $Y$ & $0.24$ 
                 & Helium mass fraction \\
    $Z_{\rm init}$ & $3.2\times 10^{-4} \ \Zsun{}$ 
                 & Initial metal mass fraction \\
    $\etaSN$ &  $0.2$ & SN ejecta mass fraction \\
    $\mSN$ & $5 \ \Msun$ & Mean SN progenitor mass \\
    $y$ & $0.075$ & SN metal yield\\
    
    \bottomrule
  \end{tabular}
  \end{center}
\end{table}

\subsection{Halo finder} \label{halofinder.sec}

For identification of haloes and galaxies, we use the \adaptahop{}
halo finder \citep{Aubert2004, Tweed2009} in the most massive
submaxima (MSM) mode. We fit a tri-axial ellipsoid to each (sub-)halo
and check that the virial theorem is satisfied within this ellipsoid,
with the center corresponding to the location of the densest
particle. If this condition is not satisfied, we iteratively decrease
its volume until we reach an inner virialized region. From the volume
of this largest ellipsoidal virialized region, we define the virial
radius $\Rvir$ and mass $\Mvir$. We have checked that our results are
similar if we instead use $\Mtwoh$ and $R_{200}$, defined to produce
a spherical over-density $200$ times the critical value. For the halo
finder, we require a minimum of $20$ particles per halo. In practice,
using the notation of \citet[][App. B]{Aubert2004} we use for the halo
finder $N_{\rm SPH} = 32$, $N_{\rm HOP} = 16$, $\rho_{\rm TH}=80$, and
$f_{\rm Poisson} = 4$. We also require that a (sub-)group of particles
has at least $20$ particles during the halo/sub-halo decomposition
step. In the end, we retain for the analysis only those haloes we
consider resolved, with $\Mvir > 300 \ \mDM$, where $\mDM$ is the DM
particle mass. For this work, we also ignore sub-haloes in our
analysis.

We also identify galaxies with \adaptahop{}. Here we require at least
$10$ stellar particles per galaxy and we add sub-groups to the main
bodies. The galaxy-finder parameters are: $N_{\rm SPH} = 10$,
$N_{\rm HOP} = 10$, $\rho_{\rm TH}=10^4$, and $f_{\rm Poisson} = 4$.

\subsection{Initial conditions}
We generate the cosmological initial conditions (ICs) with \music{}
\citep{Hahn2011}.  Our $\Lambda$CDM universe has cosmological
parameters $\Omega_{\Lambda}=0.6825$, $\Omega_{\rm m}=0.3175$,
$\Omega_{\rm b}=0.049$,
$\Hubble \equiv H_0/({\rm 100 \ km \ s^{-1} \ Mpc^{-1}})=0.6711$, and
$\sigma_8=0.83$, consistent with the Planck 2013 results
\citep{Ade2014}.  We assume hydrogen and helium mass fractions
$X=0.76$ and $Y=0.24$, respectively, and the gas is given an initial
homogeneous metal mass fraction of
$Z_{\rm init}=6.4\times 10^{-6} = 3.2\times 10^{-4} \ \Zsun{}$ (we
assume a Solar metal mass fraction of $\Zsun=0.02$ throughout this
work). This unrealistically non-pristine initial metallicity is used
to compensate for our lack of molecular hydrogen cooling channels in
the infant Universe, allowing the gas to cool below $10^4$ K, and
calibrated so that the first stars form at redshift $z\approx 15$.

\begin{table}
  \begin{center}
  \caption
  {Simulation volumes used in this work. The table columns are as
    follows, from left to right. Name: simulation name, with the first
    number denoting the volume width and the second number denoting the
    number of resolution elements (DM particles, coarse cells) per
    dimension. $\Lbox$: volume width. $\Dxmax$: physical width of
    coarsest grid cells. $\Dxmin$: physical width of finest grid
    cells. $\mDM$: mass of DM particles. $\mstar$: minimum initial
    mass of stellar particles (most particles have this initial mass,
    but some have an integer multiple of it). }
  \label{sims.tbl}
  \begin{tabular}{l|c|c|c|c|c}
    \toprule
    Name & $\Lbox$ & $\Dxmax$ & $\Dxmin$ & $\mDM$  & $\mstar$ \\
         & [cMpc]    & $(z=6)$  & $(z=6)$  & [$\Msun$] & [$\Msun$] \\
    \midrule    
    \Sphlg   & $10$ & $2.8$  kpc & $10.9$ pc & $2.5 \times 10^5$ & $10^3$ \\
    \Sphsmhr & $5$ & $1.4$  kpc & $10.9$ pc & $3.1 \times 10^4$ & $10^3$ \\
    \Sphsm   & $5$ & $2.8$  kpc & $10.9$ pc & $2.5 \times 10^5$ & $10^3$ \\
    \bottomrule
  \end{tabular}
  \end{center}
\end{table}

We use three sets of initial conditions with two volume sizes, as
listed in \Tab{sims.tbl}. Our main simulation volume, \Sphlg, has a
width of $10$ cMpc, a minimum (coarse) physical resolution of $19.6$
ckpc ($2.8$ kpc at $z=6$), and reaches a maximum cell resolution of
$76.3$ cpc ($10.9$ pc at $z=6$). It is populated by $512^3$ DM
particles with mass $\mDM = 2.6 \times 10^5 \ \Msun$ each. We assume
the limit of a resolved halo being at a mass corresponding to $300$ DM
particles.  This means we resolve haloes down to a mass of
$\Mvir \approx 7.5 \times 10^7 \Msun$, which is slightly above the
atomic cooling limit of $\Mvir \approx 3 \times 10^{7} \ \Msun$
\citep{Wise2014}.  The atomic cooling limit is important because below
it, primordial galaxies form inefficiently and likely contribute
little to reionization due to their self-destructive behaviour
\citep{Kimm2017}. To measure the possible contribution from low-mass
haloes that are above the atomic cooling limit but unresolved in the
\Sphlg{} volume (i.e. at masses
$\approx 3 \times 10^7 - 10^8 \ \Msun$), we use volume \Sphsmhr{},
with the same number of DM particles but half the width, thus giving
an eight times higher DM mass resolution, or
$\mDM = 3.1 \times 10^4 \ \Msun$, and haloes resolved down to
$\Mvir \approx 10^7 \Msun$. Statistical representation of massive
haloes is sacrificed, due to the smaller volume, to the gain of
resolving haloes down to the atomic cooling limit. This allows us to
probe the contribution of the smallest haloes to reionization. We keep
the maximum physical resolution in this volume the same as in
\Sphlg{}, i.e. $10.9$ pc at $z=6$, but the coarse resolution is
higher, following the DM resolution, or $1.4$ kpc at $z=6$.

The third simulation volume, \Sphsm{}, connects the other two volumes
by having the resolution of the larger volume, \Sphlg{}, and size and
initial conditions of the the smaller volume, \Sphsmhr{}. Hence we can
use it for resolution convergence studies, by comparing the properties
of galaxies to those in \Sphsmhr{}.

\subsubsection{Volume selection} \label{ICs.sec}

Our simulation volumes do not probe the scales at which the Universe
is homogeneous. Hence we are influenced by cosmic variance, such that
different IC realisations, with different random seeds for the
generation of density fluctuations, result in different halo mass
functions at later epochs.  This can, in turn, affect the luminosity
function and change the reionization history.

\begin{figure}
  \centering
  \hspace{-5mm}
  \includegraphics[width=0.50\textwidth]
    {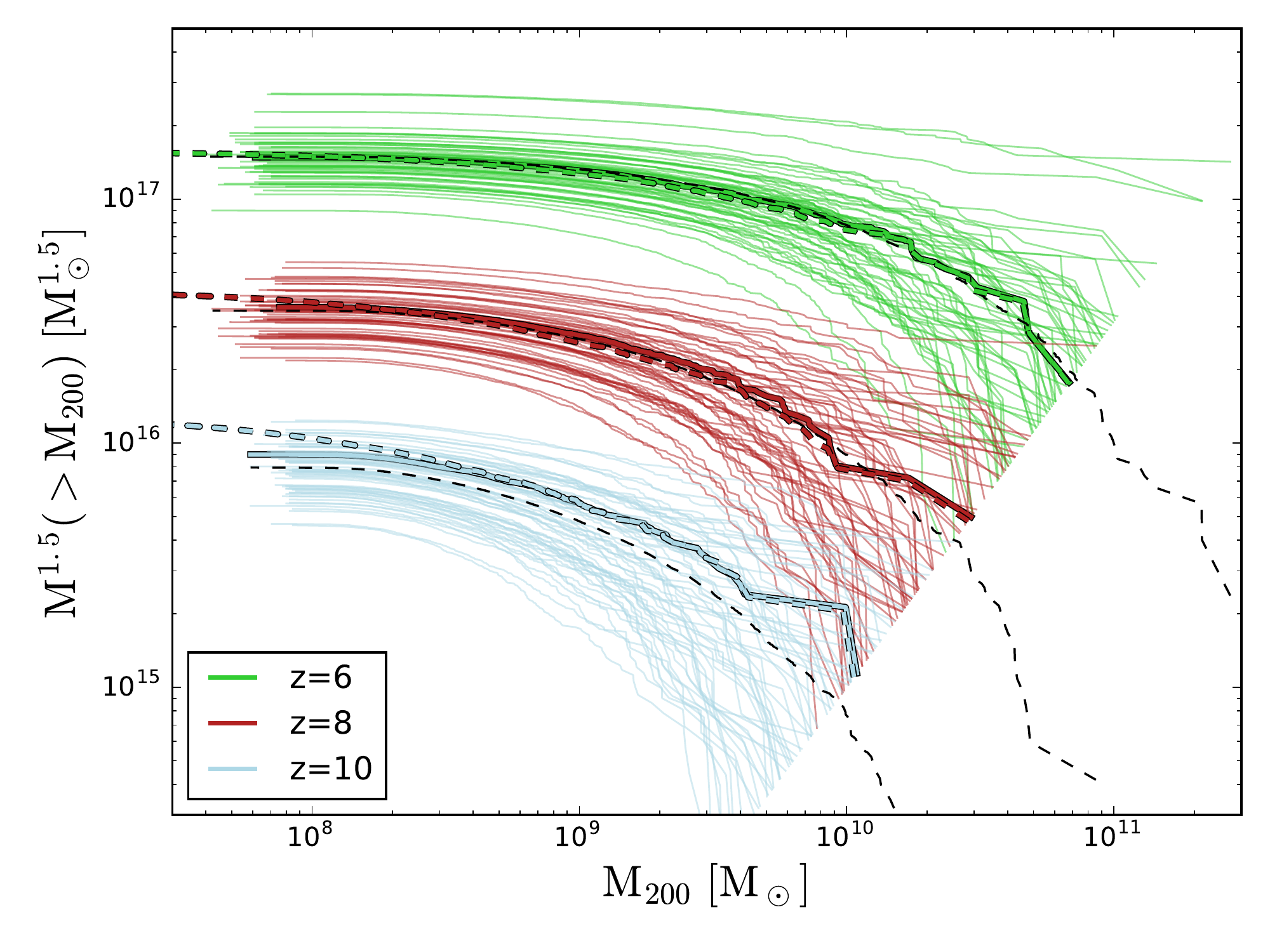}
  \caption
  {\label{ICSelection.fig}Cumulative power-$1.5$ halo mass functions
    at $z=10,8$, and $6$, as indicated in the legend, in
    $256^3$-resolution pure DM simulations run from $60$ IC
    realisations for $10$ cMpc volumes (thin coloured curves). The
    dashed black curves show the averages of all the simulations at
    the same redshifts, and the solid thick coloured curves represent
    the IC seeds we have selected for the production ($10$ cMpc)
    \sphinx{} simulations to best match the averages at all three
    redshifts. The dashed coloured curves represent the same seeds for
    a non-degraded DM mass resolution of $512^3$.}
\end{figure}

To recover a representative luminosity density and reionization
history from our limited-volume simulations, we seek to minimise the
effects of cosmic variance on the ionizing luminosity budget by
selecting the most representative ICs from a large set of
candidates. We perform $60$ pure DM simulations at a degraded
resolution of $256^3$ for both the $10$ and $5$ cMpc wide volumes,
each starting from ICs generated with a unique set of random number
seeds. We then select from those the ICs that give the most average
cumulative DM halo mass function to the power of $1.5$,
$\Mtwoh^{1.5}$, at $z=10$, $z=8$ and $z=6$. We use $\Mtwoh^{1.5}$
rather than $\Mtwoh$ because we find that it correlates better with
the total luminosity of galaxies: the ionizing luminosity scales
linearly with the star formation rate, which scales roughly as
$\Mtwoh^{1.5}$ (see \Fig{sfr_mh.fig}).

\Fig{ICSelection.fig} shows the cumulative $\Mtwoh^{1.5}$ functions at
$z=10,8$ and $6$ for the $60$ pure DM runs of the $10$ cMpc ICs in
coloured curves, and the averages of all $60$ simulations in thick
dashed black curves.  Our selected ICs are represented by thick
coloured curves, at the degraded $256^3$ DM mass resolution in solid
and the fiducial $512^3$ resolution in dashed. A comparison of the
thick coloured solid and dashed curves shows that the DM resolution
has little effect on the cumulative halo mass function (but note that
it does have an impact on the non-cumulative function at the low-mass
end, since low-mass haloes are lost when degrading the
resolution). Out of $60$ candidate ICs, it is difficult to get an
ideal average at all three redshifts. The selected ICs are very close
to the average at $z=6$ and $z=8$, but slightly top-heavy at $z=10$
and thus perhaps produce slightly too many stars at $z\gtrsim10$.

\begin{figure}
  \centering
  \includegraphics[width=0.49\textwidth]
    {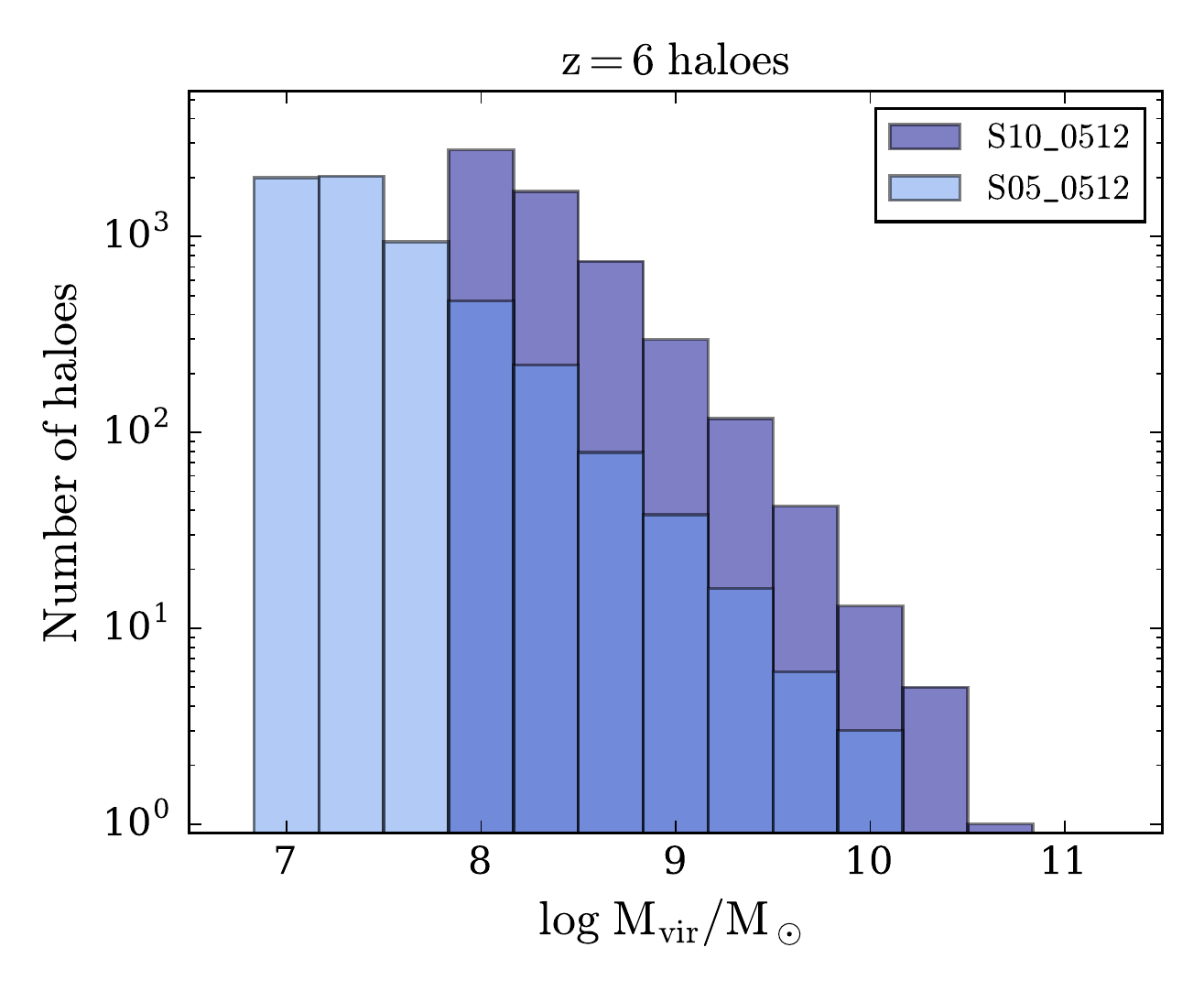}
  \caption
  {\label{halomassfunction.fig}Number of $z=6$ haloes, binned by virial
    mass ($\Mvir$), for our two main volumes; the high DM-mass
    resolution $5$ cMpc volume (light blue) and the fiducial
    resolution $10$ cMpc volume (darker blue). At this redshift, each
    volume contains approximately $5800$ haloes above the resolution
    limit of 300 DM particles (corresponding to a DM mass of
    $10^7 \ \Msun$ for the smaller volume and $8 \times 10^7 \ \Msun$
    for the larger volume).}
\end{figure}

\Fig{halomassfunction.fig} shows the number of haloes in our two
simulation volumes, binned by halo mass. At $z=6$, each volume
contains about $5800$ haloes above the resolution limit. The smaller
and larger volumes have maximum halo masses of
$\Mvir=1.1\times10^{10} \ \Msun$ and $4.7\times10^{10} \ \Msun$,
respectively, at $z=6$.

\subsection{Adaptive refinement}
In the cubical octree structure of \ramses{}, the cell refinement
level $\ell$ sets the cell width
$ \Delta x_{\ell} = 0.5^{\ell} \ \Lbox$, where $\Lbox$ is the width of
the volume. Taking our $10$ cMpc volume, the coarsest level is
$\levelmin=9$, corresponding to $2^{\ell}=512$ coarse cell widths and
a minimum resolution of $19.6$ ckpc ($2.8$ kpc at $z=6$). Starting at
this coarsest level, cells are adaptively refined to higher levels, up
to $\levelmax=17$, which corresponds to a maximum physical resolution
of $10.9$ pc at $z=6$.

\begin{table*}
  \begin{center}
  \caption
  {Photon group energy (frequency) intervals and properties. The
    energy intervals defined by the groups are indicated in units of
    eV by $\epsilon_0$ and $\epsilon_1$ and in units of \AA{}ngstrom
    by $\lambda_0$ and $\lambda_1$. The last four columns show photon
    properties derived every $10$ coarse time-steps from the stellar
    luminosity weighted SED model. These properties evolve over time
    as the stellar populations age, and the approximate variation is
    indicated in the column headers (the difference is similar in the
    two SED models we use in this work). $\eavg$ denotes the photon
    energies, while $\csavg_{\hi}$, $\csavg_{\hei}$, and
    $\csavg_{\heii}$ denote the cross sections for ionization of
    hydrogen and helium, respectively.}
  \label{groups.tbl}
  \begin{tabular}{l|R{2}{2}R{2}{2}rr|rrrr}
    \toprule
    Photon & \multicolumn{1}{c}{$\epsilon_0$ [eV]} 
    & \multicolumn{1}{c}{$\epsilon_1$ [eV]}
    & $\lambda_0$ [\AA{}] & $\lambda_1$ [\AA{}]
    & $\eavg$ [eV] & $\csavg_{\hi} \, [\cs]$ 
    & $\csavg_{\hei} \, [\cs]$ & $\csavg_{\heii} \, [\cs]$ \\
    group & \multicolumn{1}{c}{} 
    & \multicolumn{1}{c}{}
    &  & 
    & $\pm 20 \%$ & $\pm 10 \%$ 
    & $\pm 10 \%$ & $\pm 25 \%$ \\
    \midrule
    UV$_{\hi}$ & 13.6 & 24.59 & $9.1 \times 10^{2}$ & $5.0 \times 10^{2}$ & 18.3 
            & $3.2 \times 10^{-18}$ & 0 & 0 \\
    UV$_{\hei}$    & 24.59 & 54.42 & $5.0 \times 10^{2}$ 
            & $2.3 \times 10^{2}$ & 33.9 
            & $6.2 \times 10^{-19}$& $4.7 \times 10^{-18}$ & 0 \\
    UV$_{\heii}$   & 54.42  & \multicolumn{1}{c}{$\infty$} 
            & $2.3 \times 10^{2}$ & 0
            & 63.3 & $9.3 \times 10^{-20}$ & $1.4 \times 10^{-18}$ & 
                                        $1.2 \times 10^{-18}$ \\
    \bottomrule
  \end{tabular}
  \end{center}
\end{table*}

A cell is flagged for refinement into $8$ equally sized children cells
if: i) $\MDMcell + \frac{\Omegam}{\Omegab}\Mbaryonscell > 8 \ \mDM$,
where $\MDMcell$ and $\Mbaryonscell$ are the total DM and baryonic
(gas plus stars) masses in the cell, or ii) the cell width is larger
than a quarter of the local Jeans length,
\begin{align}
  \LJeans = \sqrt{\frac{\pi \csound^2}{G \rho}}
  = 16 \ {\rm pc} \ \left( \frac{T}{1 \ {\rm K}} \right)^{1/2}
  \ \left( \frac{\nh}{1 \ \cci} \right)^{-1/2},
\end{align}
where $\csound=\sqrt{\gamma \kb T / \mp}$ is the speed of sound, $G$
the gravitational constant, $\rho$ the gas mass density, and $\nh$
the hydrogen number density.

Contrary to the default refinement strategy in \ramses{}, which is to
increase the maximum refinement level at fixed scale factor intervals
to roughly maintain a constant maximum physical resolution, we keep
the maximum refinement level fixed throughout the simulation. Hence
the minimum physical cell widths are smaller at the start of the
simulation than at $z=6$. The difference in minimum cell size is not
much different than at $z=6$ though, since the maximum level is first
triggered at $z\sim20$ and thus the finest cells are $\sim3$ times
smaller (i.e. $\sim3.5$ pc) than at $z=6$.

The resolution of the gravitational force is the same as that of the
gas, with the gravitational potential calculated for each cell. In the
fiducial-resolution \Sphsm{} and \Sphlg{} runs, the DM density is
smoothed to the second-finest levels, whereas in the high-resolution
\Sphsmhr{} runs, the DM density is applied to the finest level. This
smoothing of the DM density is decided from analogue pure DM
simulations using the refinement criteria stated above but without any
limit in the maximum refinement. The maximum refinement actually
reached in those pure DM simulations sets the maximum DM density
level: $\levelmax$ for the high-resolution RHD simulations but
$\levelmax-1$ for the fiducial resolution.

\subsection{Radiation}\label{radidation_setup.sec}

The methods for photon injection, M1 moment advection, and the
interaction of radiation with hydrogen and helium gas via
photo-ionization, heating, and direct radiation pressure, are fully
described in \cite{Rosdahl2013}. The radiation is split into three
photon groups, bracketed by the ionization energies for $\hi$, $\hei$,
and $\hei$, and shown in \Tab{groups.tbl}. Radiation interacts with
hydrogen and helium via photo-ionization, heating, and momentum
transfer \citep{Rosdahl2015a}. The group properties (average energies
and cross sections to hydrogen and helium) are updated every $10$
coarse time-steps from luminosity-weighted averages of the spectra of
all stellar populations in the simulation volume, as described in
\cite{Rosdahl2013}. We do this so that at any time, the cross sections
and photon energies are representative of the average photon
population. Hence, as indicated in the \Tab{groups.tbl} (and
\Fig{SEDs.fig} in \App{SEDprops.app}), the energies and cross sections
change by a few tens of percent over the course of a simulation as the
contributions from both old and metal-rich stellar populations
increases.

In this work, we ignore the effects of radiation at sub-ionizing
energies ($< 13.6$ eV). We have run a sub-set of small-volume
simulations at the fiducial resolution with added optical and
reprocessed infrared radiation groups, and including multi-scattering
radiation pressure on dust as described in \cite{Rosdahl2015a,
  Rosdahl2015}, but found a negligible impact. This is likely because
of the low metal and dust content at the early epoch we are concerned
with.

Radiation is advected on the grid with an explicit solver, and
therefore the simulation time-step is subject to the Courant
condition. Since the radiative transfer (RT) time-step can become
significantly smaller than the hydrodynamic time-step, we subcycle the
RT on each AMR level, with a maximum of $500$ RT steps performed after
each hydro step (if the projected number of RT steps exceeds $500$,
the hydro time-step length is decreased accordingly). During the
sub-cycling on each level, radiation is prevented from propagating to
other levels and the radiation flux across boundaries is treated with
Dirichlet boundary conditions: when advecting radiation on grid
refinement level $\ell$, the radiation density and flux in
neighbouring cells at the finer ($\ell+1$) and coarser ($\ell-1$)
levels is fixed at the states from the last RT steps performed at
those levels. Radiation can cross boundaries \emph{into} the level
currently being sub-cycled, but to prevent `bursts' of radiation
across level boundaries, levels $\ell \pm1 $ are not updated when
sub-cycling level $\ell$: these updates are performed only when those
other levels are being RT sub-cycled. In this scheme, perfect photon
conservation is not strictly maintained across level boundaries,
although we have verified with tests that the number of photons is
conserved to the level of a few percent or less. For details of the
same sub-cycling scheme in the context of flux-limited diffusion, see
\cite{Commercon2014}.

To prevent prohibitively small time-steps and a large number of RT
subcycles, we use the variable speed of light approximation (VSLA)
described in \cite{Katz2017}, with a \emph{local} speed of light
$\cred = \fc c$ reduced by a factor $\fc$ from the real speed of light
$c$. Here, we apply a relatively slow speed of light in the finest
grid cells, which typically represent the ISM of galaxies, and
increase the speed of light by a factor two for every coarser level,
up to a light speed fraction $\fc=0.2$ in the coarsest cells. These
correspond to diffuse voids in the simulation volume (to be precise,
the coarsest cells all have density less than eight times the critical
density). This naturally follows the velocity increase of ionization
fronts with lower gas density \citep{Rosdahl2013} and allows us to
reduce the cost of the calculation by reducing the speed of light in
the regions that constrain the global time step.

In our $10$ cMpc volume, the speed of light fraction is set to
$\fc=[0.2, 0.1, 0.05, 0.025, 0.0125]$ for refinement levels
$\ell=[9, 10, 11, 12, 13]$, and is fixed at $\fc=0.0125$ for
$\ell\ge 13$. Even with the VSLA, using the full speed of light at the
coarsest level(s) is prohibitively expensive. However,
\cite{Katz2018} show that the reionization history is
fairly well converged with this setup \citep[see
also][]{Gnedin2016}. The main deviations are $\lesssim10\%$ overshoot
in the volume-weighted neutral fraction, $\Qhi$, during the final
stage of reionization, when voids are being ionized, and a few tens of
percent overshoot (undershoot) in the volume-weighted neutral fraction
(photo-ionization rate) when the volume is completely reionized. For
our comparison between SED models, these VLSA effects are
negligible. Note that the variable speed of light conserves the total
number of photons. The density of photons for group $i$, $\Nphot_i$
varies locally and inversely with the speed of light, but the flux,
$\cred \Nphot_i$ remains exactly constant \citep[see][]{Rosdahl2013,
  Katz2017}.

With our set-up for RT sub-cycling and the VSLA, the radiative
transfer is rarely sub-cycled more than $10$ times on any given
level. The only exception to this occurs early on in the simulations
when the first stars have just been born and are emitting ionizing
photons but SN have yet to occur.  In this case, the hydro time steps
are still very long compared to the RT time steps.
 
\subsubsection{Single and binary star radiation sources}

In this work, we compare the stellar emission from two SED models; the
\galaxev{} model of \cite{Bruzual2003}\footnote{
  \protect\url{www.bruzual.org/bc03/Updated\_version\_2016/ BC03\_basel\_kroupa.tgz}}
and the Binary Population and Spectral Synthesis code
\citep[\bpass:{}][]{Eldridge2007,
  Stanway2016}\footnote{\protect\url{http://bpass.auckland.ac.nz/2.html}}. The
latter includes binary stars while the former does not; for
simplicity, we refer to the \cite{Bruzual2003} model as singles and
the \bpass{} model as binaries throughout this paper.

For the single stars, we use the model generated with the
semi-empirical BaSeL 3.1 stellar atmosphere library
\citep{Westera2002} and a \cite{Kroupa2001} initial mass function
(IMF). For the binaries we use the available IMF closest to
\cite{Kroupa2001}, with slopes of $-1.3$ from $0.1$ to $0.5 \ \Msun{}$
and $-2.35$ from $0.5$ to $100 \ \Msun$. In \App{SEDprops.app} we
describe in detail how we extract luminosities and photon group
properties from the assumed SEDs, and show how the group properties
evolve with age and metallicity.

\begin{figure}
  \centering
  \includegraphics
    {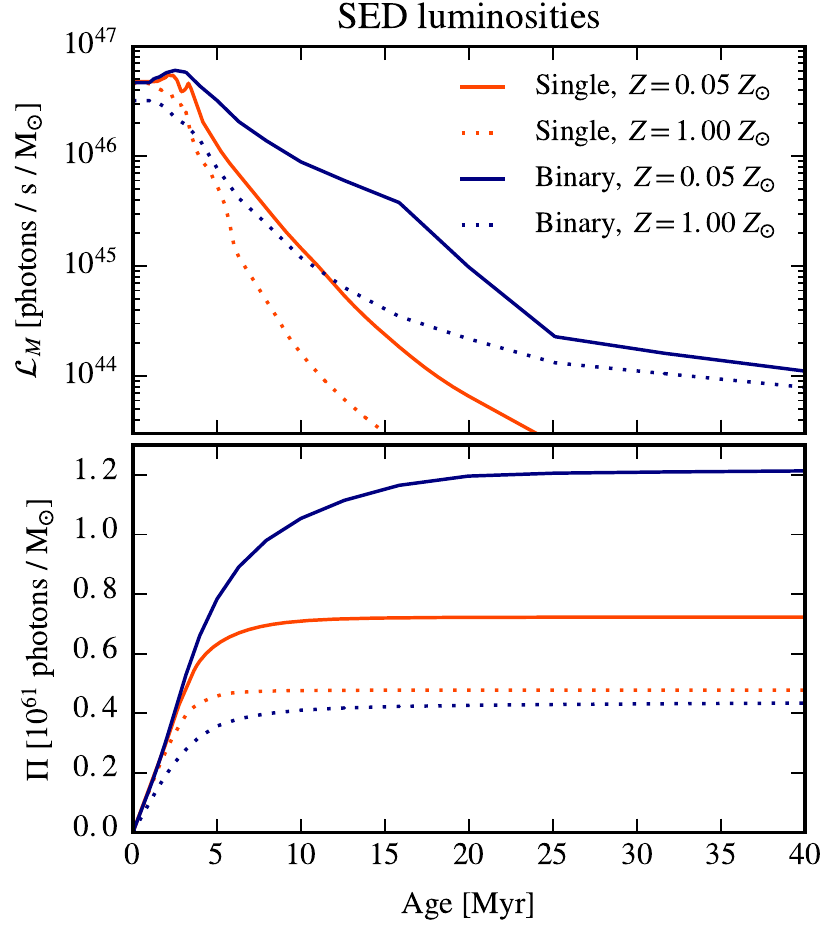}
  \caption
  {\label{SEDlums.fig}Ionizing luminosities ($\IonLumMass$; {\bf upper
      panel}) and cumulative number of hydrogen-ionizing photons
    ($\Pi$; {\bf lower panel}) as a function of stellar population age
    for SED models with single stars only \citep[]{Bruzual2003} in red
    and binary stars \citep[]{Stanway2016} in blue. The solid curves
    are with Solar metallicity and the dashed with $1/20$th Solar. At
    any metallicity, the luminosity decline is slower with binary
    stars included. At low metallicity, the inclusion of binary stars
    leads to almost twice as many ionizing photons as the case of
    single stars only.}
\end{figure}

\Fig{SEDlums.fig} shows the instantaneous and integrated luminosities
($\IonLumMass$ and $\Lumint$, respectively) for the two sets of SED models,
for Solar metallicity ($\Zsun$) and $1/20 \ \Zsun$, which is the
lowest metallicity available in \bpass{}\footnote{For the
  \cite{Bruzual2003} model, we have used the interpolation scheme in
  \ramsesrt{} to retrieve the low metallicity curve and compare with
  \bpass{}, since the exact metallicity of $1/20$th Solar is not
  tabulated in the official distribution. }. There are two important
differences to note. First, for low metallicity, the binaries model
emits almost twice as many ionizing photons than the singles
model. Second, at any metallicity, the ionizing luminosity decreases
significantly faster with stellar age for the singles model compared
to the binaries. For example, for $Z=0.05 \ \Zsun$, a singles
population emits $86\%$, $97\%$, and $99\%$ of its ionizing photons in
the first $5$, $10$, and $15$ Myr, respectively (the total is taken at
$20$ Gyr, where the model ends), while the binaries population has
emitted only $63\%$, $85\%$, and $93\%$ of all its ionizing photons at
the same ages.

Both differences may cause the binaries SED model to reionize the
Universe earlier than the singles model. However, simply increasing
the number of available photons by a small factor does not necessarily
cause significantly earlier reionization, as the radii of ionization
bubbles scale only as luminosity to the power of one-third
\citep{Stromgren1939}. The more prolonged emission with binaries is
likely more important because what matters most is how many photons
escape the dense ISM and penetrate out into the low density IGM where
only $~1$ photon per baryon is required to maintain ionization.  Due
to the interplay with stellar feedback, a stellar population is born
in dense gas where the ionizing radiation is absorbed very efficiently
in keeping up with the rapid rate of hydrogen recombinations, but as
the population gets older, feedback processes can clear away the gas,
allowing ionizing radiation to escape into the IGM \citep{Kimm2014,
  Ma2016, Kimm2017, Trebitsch2017}.

\subsubsection{Unresolved escape fractions}

In previous simulations of reionization, the escape fraction of
ionizing radiation on unresolved scales, $\fescsr$, is typically
calibrated such that the injected radiation is either higher or lower
than that given by the SED model used. A sub-unity $\fescsr$ is then
interpreted as modelling under-resolved over-densities, correcting for
underestimated recombination rates and too many photons escaping the
unresolved region, while $\fescsr>1$ models boost the luminosity of
stellar sources to compensate for e.g. under-resolved escape channels,
under-resolved turbulence \citep{Safarzadeh2016}, or secondary
ionization from X-ray binaries \citep{Shull1985}. The calibration on
$\fescsr$ is usually done in such a way that the Universe is reionized
at the roughly the correct redshift ($z\approx 10-6$, depending on the
definition of `reionized'), as determined from observations.

In this work we do not perform calibration on unresolved escape
fractions, always setting $\fescsr=1$. This does not mean that we are
sure to fully resolve the escape of ionizing radiation from sources
deep within galaxies. However, since we are primarily interested in
comparing reionization histories and escape fractions with two
different SED models, and since we find that the two SED models result
in reionization histories that bracket current observational
constraints, there is no particular need for calibration. This
nonessential calibration of $\fescsr$ can be interpreted as either a
balance between under-resolved clumps and under-resolved escape
channels, or, more likely, an indication that the escape of radiation
is governed more by large-scale feedback, captured with $\sim 10$ pc
resolution, than the detailed turbulent motions of gas on pc and
sub-pc scales.

\subsection{Gas thermochemistry}
The non-equilibrium hydrogen and helium thermochemistry, coupled with
the local ionizing radiation, is performed with the quasi-implicit
method described in \cite{Rosdahl2013} via photo-ionization,
collisional ionization, collisional excitation, recombination,
bremsstrahlung, homogeneous Compton cooling/heating off the
cosmic-microwave background, and di-electronic recombination. Along
with the temperature, and photon fluxes, we track in every cell the
ionization fractions of hydrogen, singly, and doubly ionized helium
($\xhii$, $\xheii$, $\xheiii$, respectively), and advect them with the
gas, like the metal mass fraction, as passive scalars. The
thermochemistry is operator split from the advection of gas and
radiation and performed with adaptive-time-step sub-cycling on every RT
time-step (i.e. the thermochemistry is sub-cycled within the RT, which
is sub-cycled within the hydrodynamics). To reduce the amount of
thermochemistry sub-cycling, we use the `smoothing' method for
un-splitting the advection and injection of photons from the
thermochemistry, described in \cite{Rosdahl2013}. We assume the
on-the-spot approximation, whereby we ignore recombination emission of
ionizing photons, assuming it is all absorbed locally, within the same
cell.

For $T>10^4$ K, the cooling contribution from metals is computed using
tables generated with \cloudy{} \citep[][version 6.02]{Ferland1998},
assuming photo-ionization equilibrium with a \cite{Haardt1996} UV
background.  The metal cooling is not currently modelled
self-consistently with the local radiation, which is a feature
reserved for future work.  For $T \le 10^4$ K, we use the fine
structure cooling rates from \cite{Rosen1995}, allowing the gas to
cool radiatively to a density-independent temperature floor of $15$
K. We do not apply a density-dependent pressure floor to prevent
numerical fragmentation: our star formation model (see below) is
designed to very efficiently form stars at high gas densities and, in
unison with SN and radiation feedback, prevent overly high densities
and numerical fragmentation.

\subsection{Star formation}\label{sf_model.sec}
We use the turbulent star formation (SF) criterion inspired by
\cite{Federrath2012}, as described in \cite{Kimm2017, Trebitsch2017},
and \citet[][in preparation]{Devriendt2018}.
 
The following conditions must be met locally for stars to form in a
cell:
\begin{enumerate}
\item The local hydrogen density $\nh > 10 \ \cci$ and the local
  overdensity is more than a factor $200$ greater than the
  cosmological mean (the latter is needed to prevent ubiquitous star
  formation at extremely high redshift).
\item The turbulent Jeans length is unresolved, i.e.
  $\LJeansT < \Dx$, where $\Dx$ is the finest cell width and the
  turbulent Jeans length is given by \citep{Bonazzola1987,
    Federrath2012}
\begin{align} \label{LjeansT.eq}
  \LJeansT = \frac{\pi \siggas^2 + 
                \sqrt{36 \pi \csound^2 G \Dx^2 \rho + \pi^2 \siggas^4}}
                  {6 G \rho \Dx},
\end{align}
where $\siggas$ is the gas velocity dispersion over neighbour cells
sharing vertices with the host.
\item The gas is locally convergent and at a local density maximum
  compared to the six next-neighbour cells.
\end{enumerate}

Gas that satisfies the above criteria is converted into stellar
population particles at a rate
\begin{align} \label{sf_recipe.eq}
\dot \rho_{*} = \sfeff \rho / t_{\rm ff},
\end{align}
where $\tff = \left[ 3 \pi/(32 G \rho) \right]^{1/2}$ is the local
free-fall time and $\sfeff$ is the star formation efficiency. The cell
gas is stochastically converted into collisionless stellar particles
by sampling the Poisson probability distribution for gas to star
conversion over the time-step \citep[see][for details]{Rasera2006},
such that \emph{on average}, the conversion rate follows
\Eq{sf_recipe.eq}. The initial mass of each stellar particle is an
integer multiple of $\mstar=10^3 \ \Msun$, but capped so that no more
than $90 \%$ of the gas is removed from the cell. The main distinction
of this turbulent star formation recipe from traditional star
formation in \ramses{} \citep{Rasera2006}, where $\sfeff$ is a global
(typically small) constant, is that $\sfeff$ varies locally with the
thermo-turbulent properties of the gas \citep[see][for
details]{Kimm2017, Trebitsch2017, Devriendt2018}. The local star
formation efficiency can approach and even exceed $100 \%$ (with
$\sfeff>1$ meaning that stars are formed faster than in a free-fall
time). This gives rise to a bursty mode of star formation, whereas the
classical constant $\sfeff$ recipe leads to star formation that is
more smoothly distributed in both space and time.

We modify the SF recipe as described by \cite{Kimm2017} by subtracting
rotational velocities and symmetric divergence from the turbulent
velocity dispersion in \Eq{LjeansT.eq} and the expression leading to
$\sfeff$, such that the turbulence represents only anisotropic and
unordered motion.  This leads to stronger instantaneous star formation
at the centers of galaxies and rotating clouds, but also stronger
feedback episodes, with the net effect of reducing the overall star
formation.

\subsection{Supernova feedback} \label{models.sec}

We apply individual type II SN explosions of $10^{51}$ erg by
stochastically sampling the delay-time distribution for the IMF over
the first $50$ Myr of the lifetime of each stellar particle.

We use the mechanical SN model for momentum injection from
\citet[][see also \citealt{Kimm2016b, Rosdahl2017, Kimm2017,
  Trebitsch2017}, and a similar model described in
\citealt{Hopkins2014}]{Kimm2015}. The key feature of the mechanical
feedback model is its ability to capture the correct final snow-plow
momentum of a SN remnant, by injecting it directly if the previous
adiabatic phase is not resolved, or otherwise letting it evolve
naturally. We review here the main features of the algorithm.

For one SN explosion, its radial momentum, along with the SN ejecta
mass ($\mSN$), as well as gas mass contained in the host SN cell, is
shared between the host and all neighbour cells sharing vertices with
it (corners included; thermal energy is injected into the host instead
of momentum). The amount of momentum given to each neighbour depends
on the mass loading $\chi=\Delta \mW/(\fnbor \mSN)$. Here, the
denominator is the SN ejecta mass received by the neighbour, with
$\fnbor$ a geometrical factor determining the share of SN energy and
mass that the neighbour receives (totalling to unity for each
explosion). The numerator, $\Delta \mW$, contains all the mass in the
same neighbour after the wind injection (i.e. the mass already in the
cell plus the SN remnant plus $\fnbor$ times the mass originally in
the SN host cell).

For low $\chi$, the momentum injection follows the adiabatic phase
with
\begin{align}
  \Delta p = \fnbor \sqrt{2 \chi \mSN \fe \eSN}. 
\end{align}
Here, $\fe$ is a function ensuring a smooth transition to the maximum
momentum injection for the neighbour,
\begin{align} \label{dpmax.eq}
  \Dpmax = 3 \! \times \! 10^{5} \, \Msun \, \frac{\rm km}{\rm s} \, 
    \eOne^{\frac{16}{17}} \, n_0^{-\frac{2}{17}} \ Z'^{-0.14} \ \fnbor, 
\end{align}
where $\eOne$ is the SN energy in units of $10^{51}$ ergs and always
unity in these simulations, $n_0$ is the local hydrogen number density
in units of $1 \ \cci$, and $Z'=\max \left(Z/\Zsun,0.01\right)$. This
expression for the density- and metallicity-dependent upper limit for
the snowplow momentum comes from the numerical experiments of
\cite{Blondin1998} and \cite{Thornton1998} of explosions in a
homogeneous medium and has been confirmed more recently by
e.g. \cite{Kim2015} and \cite{Martizzi2015}.

As in \cite{Kimm2017}, we increase the maximum injected momentum in
cells where ionized ($\hii$) regions are unresolved. At each SN
injection, we compare the local \stromgren{} radius $\rS$
\citep{Stromgren1939} with the cell width $\Delta x$.  Based on a
simple fit to the results of \cite{Geen2015b}, for $\rS>\Delta x$, the
magnitude of the injected momentum gradually shifts to
\begin{align} \label{dpmax_geen.eq} 
  \Dpmax^{\rm PH} = 4.2 \! \times \!
  10^{5} \, \Msun \, \frac{\rm km}{\rm s} \, \eOne^{\frac{16}{17}} \
  Z'^{-0.14} \ \fnbor,
\end{align}
due to the unresolved clearing of dense gas via photo-ionization
heating.

We assume a \cite{Kroupa2001} initial mass function (IMF), where a
mass fraction $\etaSN=0.2$ of the initial stellar population mass
explodes as SNe and recycles back into the ISM with a metal yield of
$y=0.075$ (i.e. $7.5 \%$ of the ejecta hydrogen plus helium mass is
converted to heavy elements). Integration of the Kroupa IMF to
$100 \ \Msun$ gives an average SN progenitor mass of
$\mSN=20 \ \Msun$, resulting in $\etaSN/\mSN=1$ SN explosion per $100$
Solar masses. However, we calibrate the SN feedback in order to
roughly reproduce the early Universe stellar mass to halo mass
relation (SMHM), star formation rate (SFR) versus halo mass, and UV
luminosity function (these plots are shown in \Sec{galprops.sec} and
the effects of the calibration are detailed in
\App{SNcalibration.app}). As a result, the SN rate is boosted by
assuming $\mSN=5 \ \Msun$, which gives four SN explosions per $100$
Solar masses, four times higher than the Kroupa IMF.

The boost in the rate of SN explosions is pure calibration, but can be
viewed as representing uncertain factors of: i) the high mass end of
the IMF (a less conservative integration over $6-150 \ \Msun$ gives a
SN rate close to ours), ii) numerical overcooling
\citep{Katz1996}\footnote{The \cite{Kimm2015} sub-grid model
  circumvents numerical overcooling of SN blasts by design, but even
  so overcooling may still be an issue, e.g. due to under-resolved gas
  porosity \citep{Kimm2015}.}, and iii) complementary feedback
processes such as cosmic rays \citep[e.g.][]{Booth2013, Hanasz2013,
  Girichidis2016, Pakmor2016}, stellar winds
\citep[e.g.][]{Gatto2016}, and radiation pressure in the Lyman-alpha
\citep[][]{Dijkstra2009a, Smith2016, Kimm2018} and infrared
\citep[e.g.][but see \citealt{Rosdahl2015}]{Hopkins2012c,
  Thompson2005}.

We make no SED-dependent adjustments to the SN feedback. In reality
the rate of SN explosions is inherently tied to the SED model: for
example, mass transfer between binary companions leads to an increased
number of type II SN explosions for a stellar population, and at later
times. However, to minimise the number of variables in our comparison
between SED models, we ignore these factors in the present study.

\section{Results}\label{results.sec}

\begin{figure*}
  \centering
  \includegraphics[width=1.0\textwidth]
    {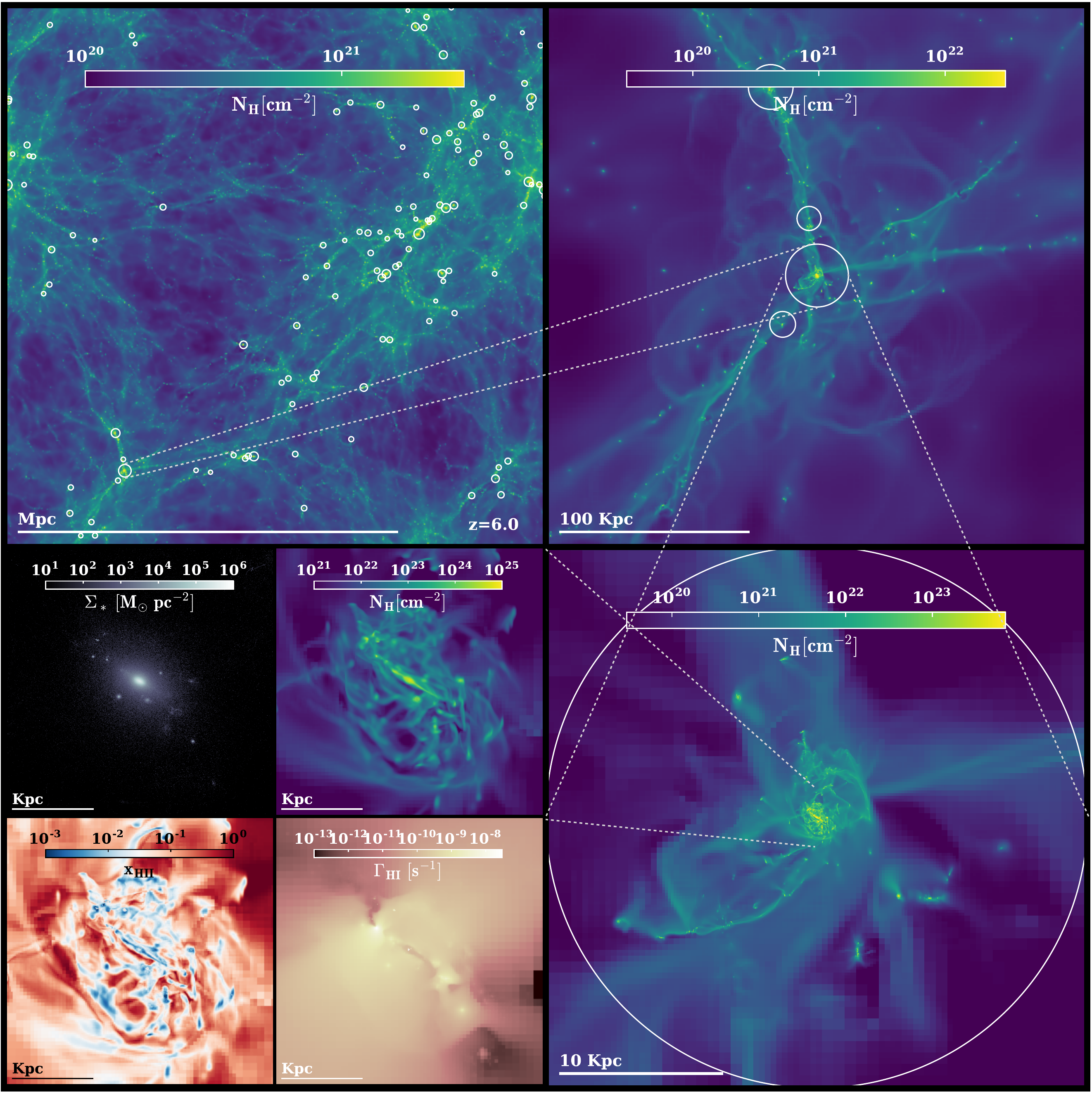}
  \caption
  {\label{map_fullbox.fig}Projections from the 10 cMpc volume at
    $z=6$, run with binary stars (\Sphlgbp). Circles mark the virial
    radii of all haloes more massive than $3 \times 10^9 \ \Msun$.
    Clockwise from top left, the first three panels show hydrogen
    column densities, $\Nh$, first for the full volume, then in the
    environment around the most massive halo
    ($\Mvir=4.7\times10^{10} \ \Msun$), and finally inside that
    halo. The four sub-panels in the bottom left corner zoom in on the
    central galaxy and show, clockwise from top right, the column
    density of gas, averaged hydrogen photo-ionization rate,
    $\Gammahi$, average density-weighted ionized hydrogen fraction,
    $\xhii$, and stellar column density, $\Sigstar$.}
\end{figure*}

We first demonstrate the range of scales captured in our cosmological
simulations.  In \Fig{map_fullbox.fig} we show maps of hydrogen column
densities ($\Nh$) in the $10$ cMpc volume (run with binary stars) at
$z=6$, starting in the top left panel with a projection of the full
volume, then clockwise showing the large-scale environment of the most
massive halo, and a zoom-in on the halo gas. The insets in the bottom
left panel show zoom-ins on the central galaxy of $\Nh$, stellar
column density ($\Sigma_*$), hydrogen photo-ionization rate
($\Gammahi$), and ionized hydrogen fraction ($\xhii$). The images show
clear signatures of violent feedback events and outflows, mergers, and
accretion, with expanding remnants of SN super-bubbles (top right
panel of halo environment) and a generally unsettled morphology of
gas.

The photo-ionization rate is a measure for the flux of ionizing
radiation and is calculated as
\begin{align}
\Gammahi=\sum_{i=1}^{N_{\rm groups}} \cred \Nphot_i \csavg_{i,\hi},
\end{align}
where $\cred$ is the local reduced speed of light, $N_{\rm groups}$ is
the number of radiation groups, $\Nphot_i$ is the photon number
density for radiation group $i$, and $\csavg_{i,\hi}$ is the hydrogen
photo-ionization cross section for group $i$.  The map of $\Gammahi$ in
\Fig{map_fullbox.fig} shows that the ionizing radiation emitted by
some stellar populations is efficiently absorbed by the ISM, for
example inside the central bulge where there is a bright spot of high
photo-ionization rate, at the location of a young cluster of stars,
surrounded by a darker region of relatively low $\Gammahi$ and high
neutral gas density. In other regions, for example north-east from the
central bulge, the radiation propagates much more freely through, and
away from, the ISM.

We now compare our galaxy populations to observables of the
high-redshift Universe and demonstrate that we reproduce an accurate
luminosity budget for our simulated patches of the Universe. We will
then present our simulated reionization histories, and finish by
comparing the redshift-evolution of escape fractions with single and
binary SED models.

\subsection{Galaxy properties}\label{galprops.sec}

\begin{figure}
  \centering
  \includegraphics[width=0.5\textwidth]
    {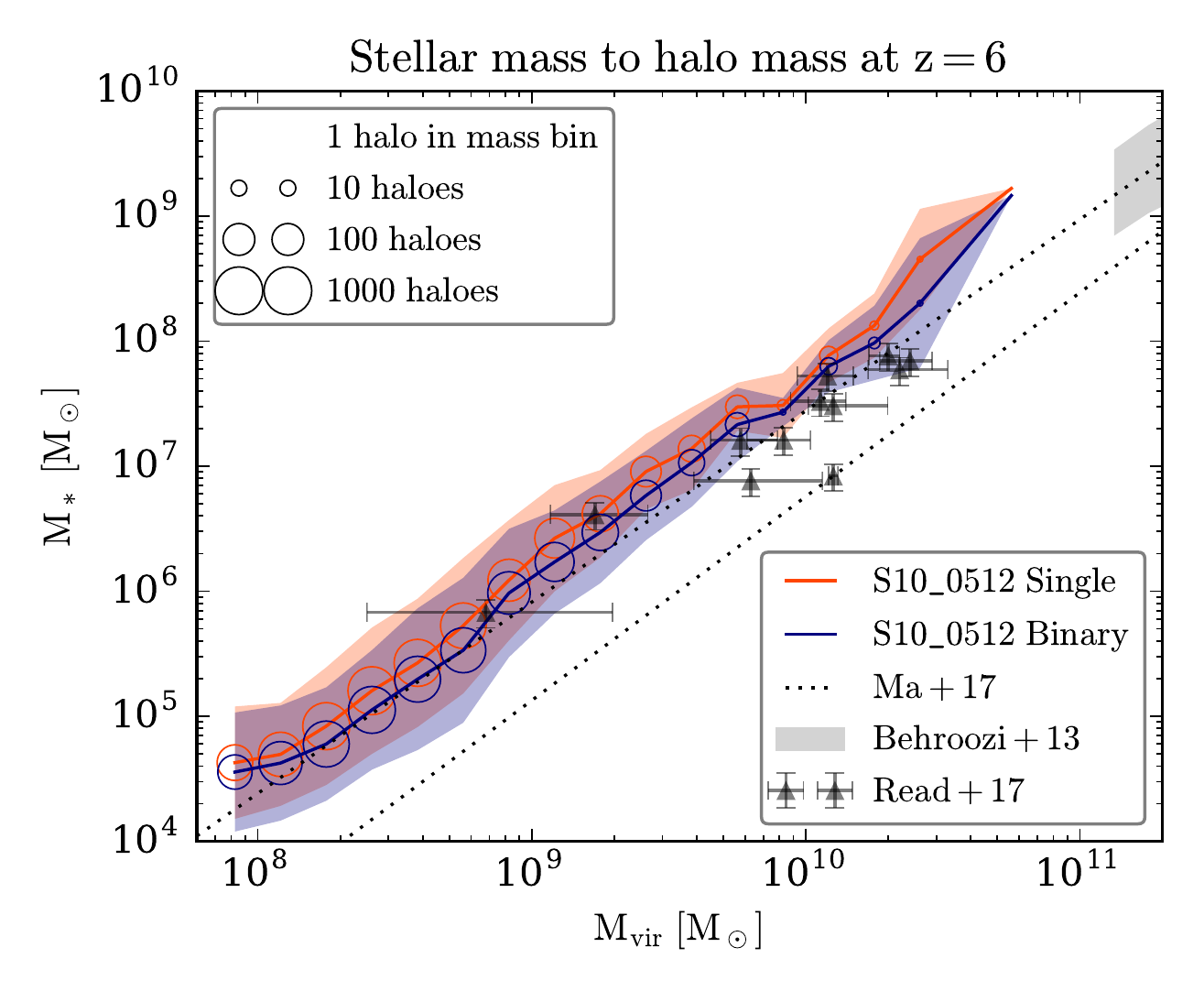}
  \caption
  {\label{smhm.fig}Stellar mass to halo mass relation in out $10$ cMpc
    volume at $z=6$. The red and blue curves show average stellar
    masses per halo mass bins for the single and binary SED models,
    respectively, with circle sizes representing the number of
    galaxies in each bin, as indicated in the top left legend, and
    shaded regions marking the standard deviation. The grey shaded
    area in the top right corner shows observational $z=6$ constraints
    from abundance matching \citep{Behroozi2013} and black triangles
    show observational constraints of $z=0$ isolated dwarf galaxies
    \citep{Read2017}. The dashed curves represent the $1-\sigma{}$
    dispersion in the simulations of \citet{Ma2017}.  The simulated
    galaxies are in fair agreement with the \citet{Ma2017} simulations
    as well as $z=0$ observations.}
\end{figure}

We begin with the $z=6$ stellar mass to halo mass (SMHM) relation. We
assign to each halo the most massive galaxy within $0.3 \ \Rvir$ (as
identified by \adaptahop). The resulting SMHM relation is shown in
\Fig{smhm.fig} for the $10$ cMpc volumes with binary and single SED
models. We have binned the haloes by mass and plotted the mean stellar
mass per bin, with the sizes of circles indicating the number of
galaxies per bin, and shaded areas showing the standard deviation. We
note that our $5$ cMpc volume results are very similar to those for
the $10$ cMpc one, only extending to lower maximum (and minimum) halo
masses.

The different SED models produce similar SMHM relations.  This is due
to the fact that, except for the SED models, we use the exact same
parameters and SN feedback model in those runs, underlining that
stellar mass is mostly regulated by SNe in our simulations. There is
secondary feedback via photo-ionization heating \citep{Rosdahl2015}
which leads to the SMHM relation in \Fig{smhm.fig} being shifted down
by $\sim25\%$ with the binary stars SED model compared to the singles
case. An identical run without any ionizing radiation (not shown)
results in an SMHM relation which is shifted up by $\sim 40 \%$
compared to the single stars SED model. We stress though that the
photo-ionization feedback is sub-dominant compared to SNe, which have
a much more dramatic effect on the stellar mass (see
\App{SNcalibration.app} and in particular \Fig{smhm_SNboost.fig} for
the effect of SN feedback).

We compare our SMHM relation in \Fig{smhm.fig} to to estimates from
rotation curves of $z=0$ dwarf galaxies \citep[][for lack of direct
observational constraints with such low-mass haloes at high
redshift]{Read2017}, with which our simulations are in good agreement,
and $z=6$ observational constraints from abundance matching
\citep{Behroozi2013} which, while not overlapping with our mass range,
are not in obvious conflict with our results. The figure also includes
predictions from the cosmological EoR simulations of \cite{Ma2017},
with dashed lines marking their $1-\sigma{}$ scatter in galaxy
masses. Our galaxy masses are slightly but systematically higher than
theirs, which is very likely due to less efficient stellar feedback,
but the SMHM relation is very similar in slope. Our scatter in galaxy
mass is similar to theirs at all masses: much like them we find that
the scatter in stellar masses decreases somewhat towards the high-mass
end, but for halo masses $\gtrsim 10^{10} \ \Msun$, our relative lack
of scatter is merely a result of limited statistics, with only a few
haloes populating this regime (see \Fig{halomassfunction.fig}).

\begin{figure}
  \centering
  \includegraphics[width=0.5\textwidth]
    {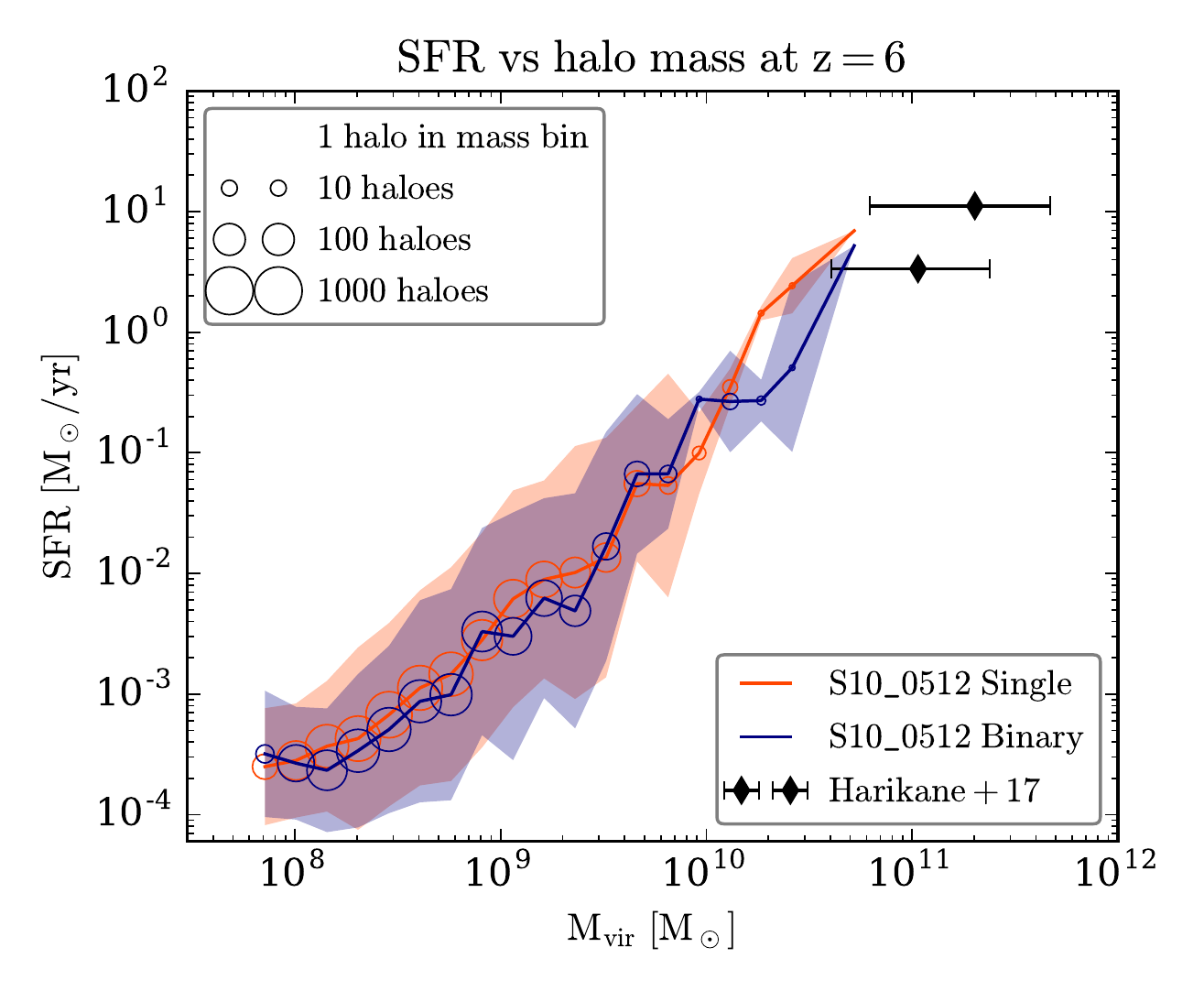}
  \caption
  {\label{sfr_mh.fig}Star formation rate (SFR), averaged over the last
    $100$ Myr, versus halo mass for our $10$ cMpc simulations with
    single (red) and binary (blue) SED models. The curves indicate
    mean SFRs in each halo mass bin and the shaded regions show the
    standard deviation. Circle sizes represent the number of galaxies
    per halo mass bin. Diamonds show observational $z=6$ estimates
    from \citet{Harikane2017}. The simulated galaxies show a large
    scatter in SFRs and a fair agreement with the observational
    estimates, although the overlap is limited.}
\end{figure}

In \Fig{sfr_mh.fig} we show the mean star formation rates
averaged over the past $100$ Myr as a function of halo mass at
$z=6$. Again there is little overlap with $z=6$ observations except at
the very highest masses of our simulated haloes. For comparison we
include estimates from \cite{Harikane2017}, with which our
highest-mass galaxies are in reasonable agreement. Again there is
slightly but systematically lower star formation efficiency with the
binary SED, hinting at non-negligible effects from the more luminous
and prolonged radiation emitted from binary stellar populations,
compared with the single stars SED.

To generate luminosity functions from our simulation outputs, we sum
the $1500$ \AA{} luminosities of all stellar particles in each galaxy
(as identified by \adaptahop). The luminosity of each stellar particle
is calculated by interpolating tables of metallicity- and
age-dependent $1500$ \AA{} luminosities for each SED. We transform the
total $1500$ \AA{} luminosity, $\LumFifteen$, of each galaxy to an
absolute AB magnitude \citep{Oke1983},
\begin{align}
  M_{\rm AB}&=-48.6 - 2.5 \log\left(
              \frac{\LumFifteen/[\mathrm{erg \ s^{-1} \ Hz]}}
              {4 \pi \ [10 \ \mathrm{pc/cm}]^2}
              \right)\nonumber \\
  &=51.595 - 2.5 \log\left(
              \frac{\LumFifteen}{\mathrm{erg \ s^{-1} \ Hz}}\right).
\end{align}
We ignore dust absorption, which is predicted by \cite{Ma2017} to
become important at $z\sim6$ for intrinsic magnitudes $\Mab\sim-20$
and brighter.

\begin{figure}
  \centering
  \includegraphics[width=0.5\textwidth]
    {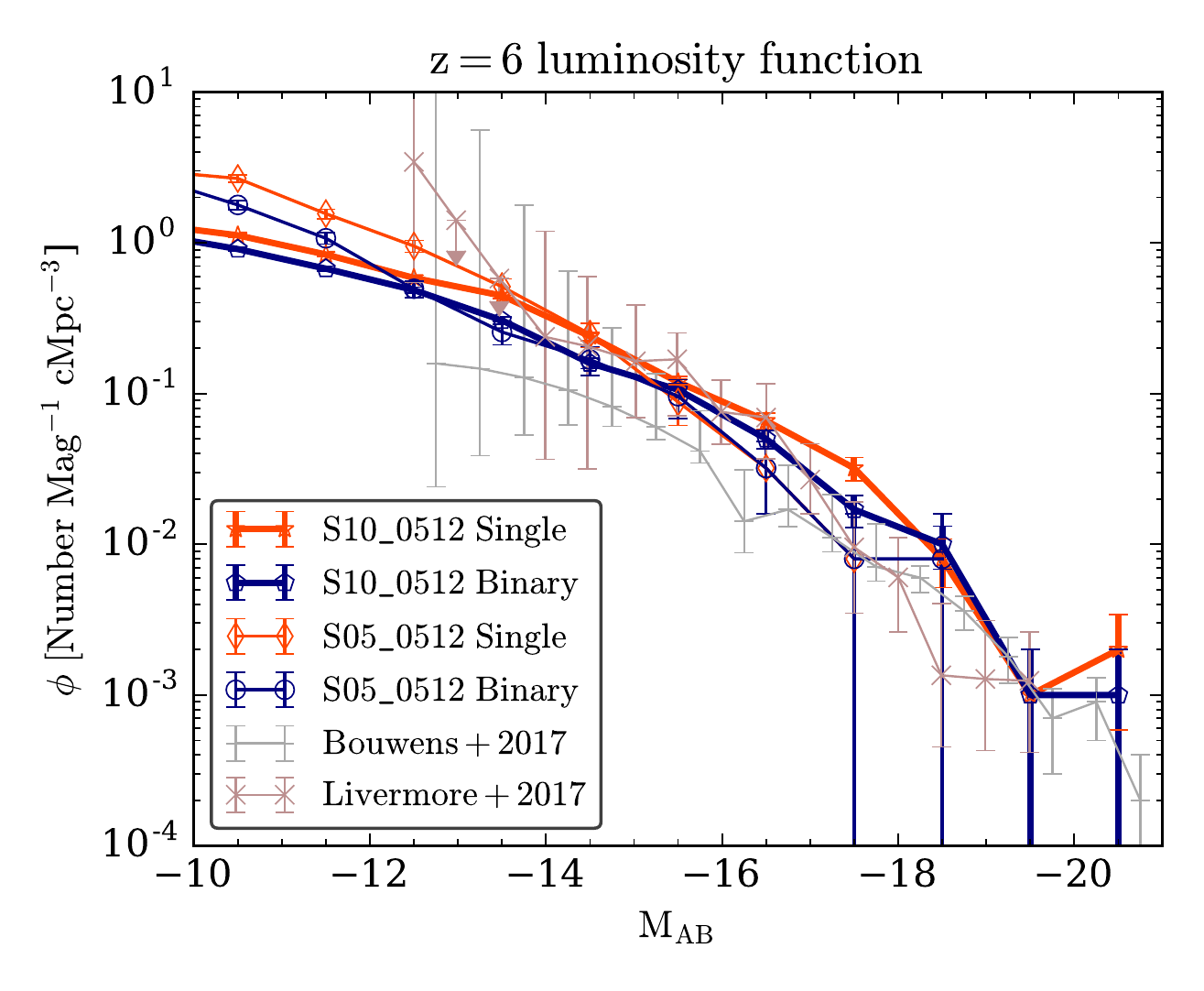}
  \caption
  {\label{flum.fig}Redshift 6 luminosity functions from our $10$ and
    higher-resolution $5$ cMpc volumes (ignoring absorption), and from
    the observational compilations of \citet{Bouwens2017} and
    \citet{Livermore2016}, as indicated in the legend. We use
    Poissonian error-bars for the \sphinx{} data, but note that the
    error from cosmic variance is typically larger. Our simulations
    agree well with the observational limits. Systematically more
    galaxies appear at the low luminosity end with the single stars
    SED.}
\end{figure}

The resulting luminosity functions are plotted in \Fig{flum.fig} and
compared to observations from \cite{Bouwens2017} and
\cite{Livermore2016}. The overall agreement with observations is good,
indicating that we produce a realistic number of photons over the
range of halo masses captured in our simulations. Thus, our results
are not compromised by an over- or under-abundance of galaxies at a
given luminosity. We note that including an absorption of
$\approx 0.5$ mag at the bright end (as suggested e.g. by
\citealt{Ma2017} or \citealt{Bouwens2015}) would further improve the
agreement of our results with observations.  It should also be noted
that our statistics at the luminous end are very limited, with the
most luminous bins in each of our volumes containing only $1-2$
galaxies. For future comparisons, we tabulate our predicted luminosity
functions with the binary SED, at $z=10,8,$ and $6,$ in
\App{flum.app}.

At the faint end, for $\Mab\gtrsim-13$, discrepancies appear for the
box sizes, with a systematically larger number of faint galaxies for the
$5$ cMpc volume than for the $10$ cMpc volume (this is true both for
the single and binary SED models, though the discrepancy appears in
slightly brighter galaxies with single stars). The discrepancies are
due to the different cosmological initial conditions (and not to
differences in resolution: luminosity functions for the $5$ cMpc box
with varying resolution are much better converged, as shown in
\App{res_conv.app}).

Overall, due to strong calibrated SN feedback, our simulated volumes
contain galaxies that are in good agreement with observations in terms
of stellar masses, star formation rates, and luminosity
functions. Hence, the simulated production of ionizing photons per
volume should roughly match the real production of photons over the
range of halo masses represented in our simulations, assuming the SED
models are accurate.

\subsection{Reionization history} \label{Reionization.sec}
\begin{figure*}
  \centering
  \subfloat
  {\includegraphics[width=0.9\textwidth]
    {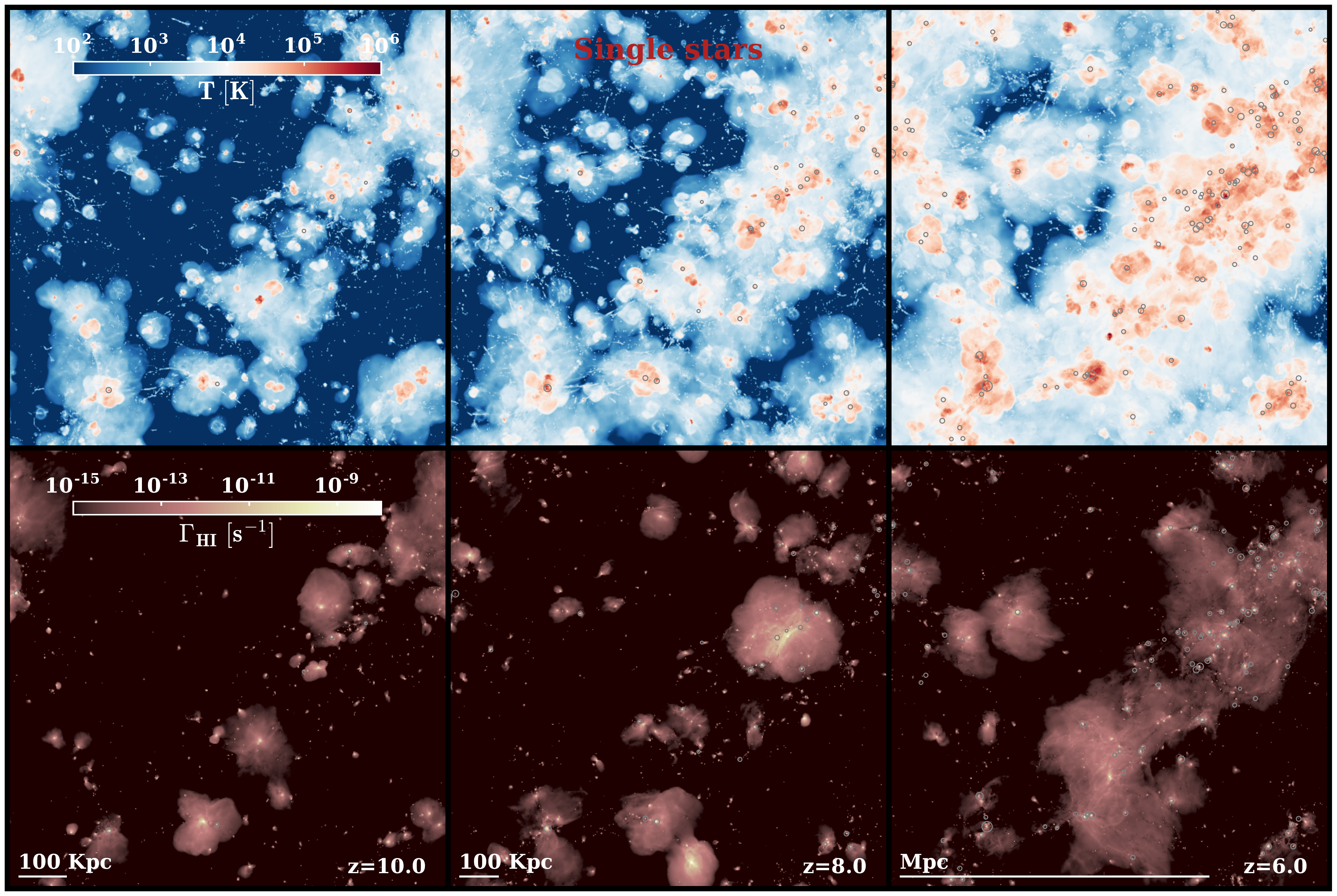}}
  \vspace{-0.2mm}
  \subfloat
  {\includegraphics[width=0.9\textwidth]
    {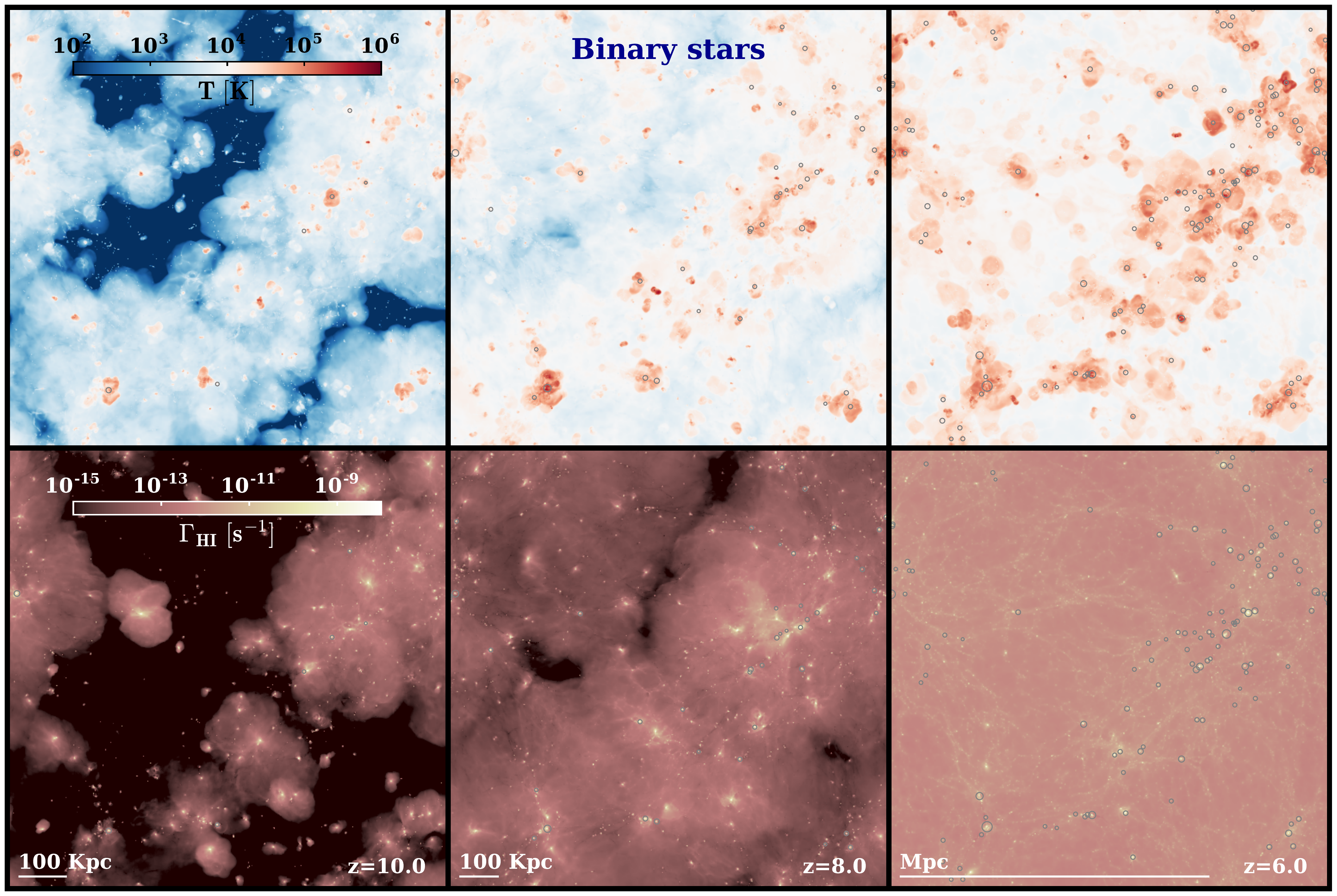}}
  \caption
  {\label{map_reionization.fig}Reionization of the $10$ cMpc volume
    with single and binary SED models (upper and lower set of panels,
    respectively). Each set of panels shows full volume mass-weighted
    projections of temperature (upper rows) and photo-ionization rate
    $\Gammahi$ (lower rows) at decreasing redshift as indicated in the
    $\Gammahi$-maps. The circles mark all haloes more massive than
    $3 \times 10^9 \ \Msun$. In the temperature maps, the blue colour
    shows cold neutral gas, white to red-ish depicts warm
    photo-ionized gas, and darker red indicates gas shock-heated by
    galactic winds. By comparing the upper and lower set of panels, it
    is clear that the inclusion of binary stars (in the lower set of
    panels) hastens the reionization process. }
\end{figure*}

\begin{figure*}
  \centering
  \includegraphics[width=0.8\textwidth]
    {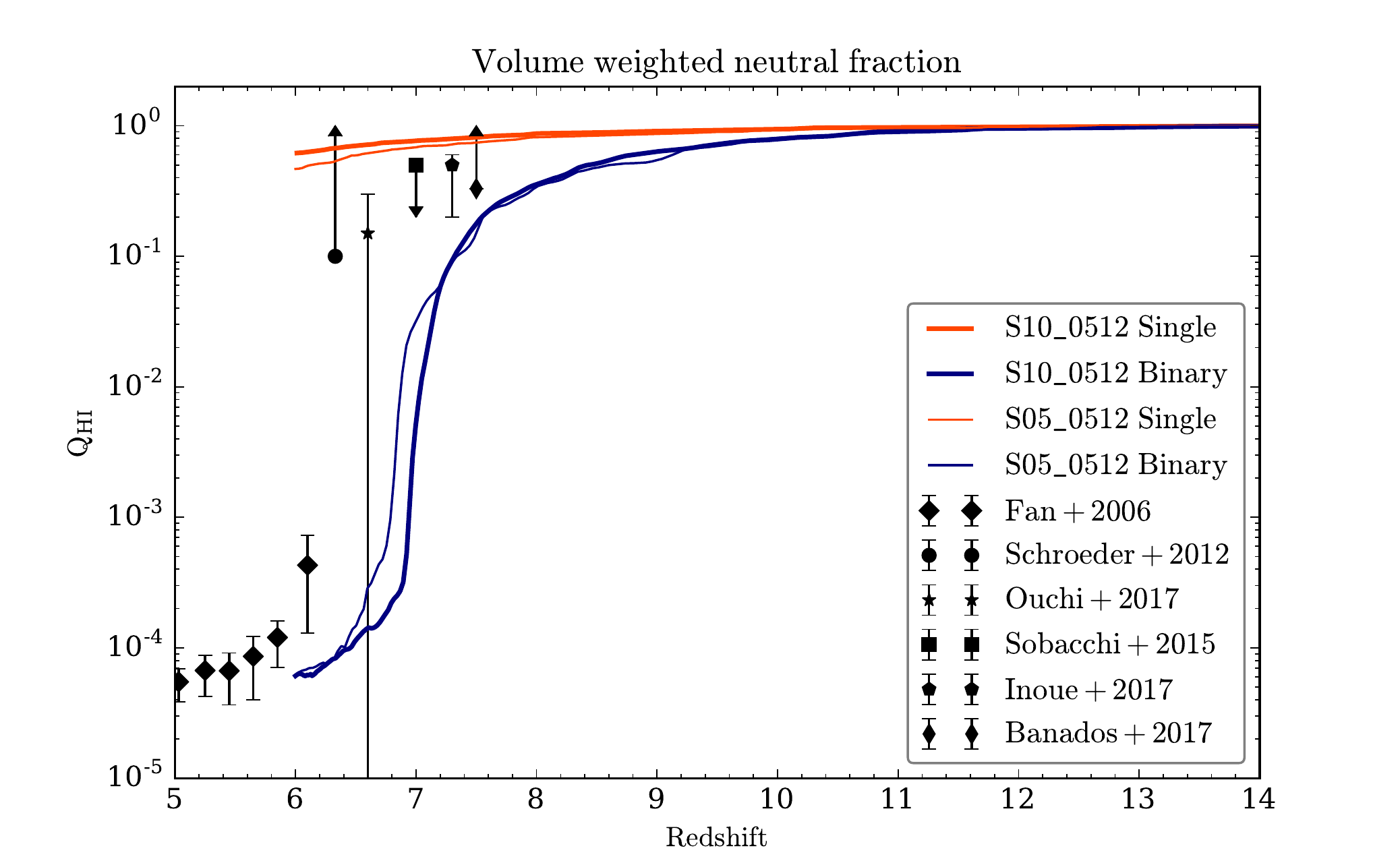}
  \caption
  {\label{xhi.fig}Evolution with redshift of the volume-filling
    neutral fraction, $\Qhi$. Our $10$ and (higher DM-mass resolution)
    $5$ cMpc volumes are shown, each run with both single and binary
    SED models, as indicated in the legend. For comparison we show
    model-dependent observational estimates from \citet{Fan2006a},
    \citet{Schroeder2012}, \citet{Ouchi2017}, \citet{Sobacchi2015},
    \citet{Inoue2017}, and \citet{Banados2017}, as indicated in the
    legend. Changing the volume size and/or resolution has very little
    effect on the reionization history, whereas changing to a SED that
    includes binary stars dramatically hastens reionization.}
\end{figure*}

We now turn to the reionization histories produced in our volumes with
single and binary SED models. \Fig{map_reionization.fig} shows a
visual comparison of the volume for the two SED models, with single
stars in the upper set of panels and binaries in the lower ones.  The
maps show projected distributions through the simulated volume of
density-weighted gas temperature $T$ (upper rows) and density-weighted
hydrogen photo-ionization rate $\Gammahi$ (lower rows) at redshifts
$10, 8$, and $6$ (from left to right). The qualitative extent of
ionized regions can be judged from the temperature projections, where
deep blue marks cold neutral regions, light-blue to light-red shows
photo-ionized gas, and darker red indicates gas that has been heated
by SNe and, to a lesser extent, virial shocks. Comparing the upper and
lower set of panels, it is clear that the IGM is ionized sooner and to
a greater extent with binary stars included. At all redshifts shown,
the ionized regions are much larger with binary stars and at $z=6$ the
individual \hii{} bubbles have merged. Without binaries there is still
a significant fraction of the volume which remains neutral and with
photo-ionization rates far below the $z\approx6$ observational
estimates of a few times $10^{-12} \ \sm$ \citep[][]{Calverley2010,
  Wyithe2010, DAloisio2018}.

\Fig{xhi.fig} shows the redshift evolution of the volume-filling
fraction of neutral hydrogen, $\Qhi$, for the $10$ and $5$ cMpc wide
volumes, with single and binary SED models.  Black datapoints in the
same plot show model-dependent observational estimates from
\citet{Fan2006a}, \citet{Schroeder2012}, \citet{Ouchi2017},
\citet{Sobacchi2015}, \citet{Inoue2017}, and \citet{Banados2017}. The
single and binary SED models produce very different reionization
histories. The volumes are only $\sim 50 \%$ reionized at $z=6$ with
the single stars SED, while they are completely reionized by
$z\approx7$ with binaries included. Note that for each SED, the two
different curves are not only for different volume sizes (and hence a
different range of halo masses) but also for different DM mass
resolution, so the results are robust to both these factors. In
\App{res_conv.app} we also show that these reionization histories are
robust with respect to resolution only (i.e. not changing the volume
size at the same time). This insensitivity to box size is in part due
to our method for selecting initial conditions to minimise cosmic
variance effects (see \Sec{ICs.sec}). It is also an indication that
our volumes are mostly ionized by intermediate-mass haloes, i.e. in the
mass range $\Mvir\approx10^8-10^{10} \ \Msun$, since this is the mass
range overlapping our two volumes (see \Fig{halomassfunction.fig}).

Neither SED model produces a reionization history in perfect agreement
with the observational limits, but it is compelling and reassuring
that variations in state-of-the-art SED modelling produce reionization
histories that bracket the observational limits, without any
calibration in unresolved escape fractions.

Note that in order to produce the good agreement with observations of
the stellar mass to halo mass, SFR to halo mass, and luminosity
function, we artificially boost the rate of SN explosions by a factor
of four compared to that derived from a \cite{Kroupa2001} IMF. Without
the factor four boost, our stellar masses and SFRs are systematically
higher than observations by factors of a few, as shown in
\App{SNcalibration.app}. Likewise, the resulting luminosity function
is far too shallow compared to observations. However, as we also show
in \App{SNcalibration.app}, our main results are surprisingly
insensitive to the feedback calibration, with remarkably similar
reionization histories produced with and without the SN boost, even if
galaxy luminosities are vastly overestimated in the un-boosted SN
case.

\begin{figure}
  \centering
  \includegraphics[width=0.5\textwidth]
    {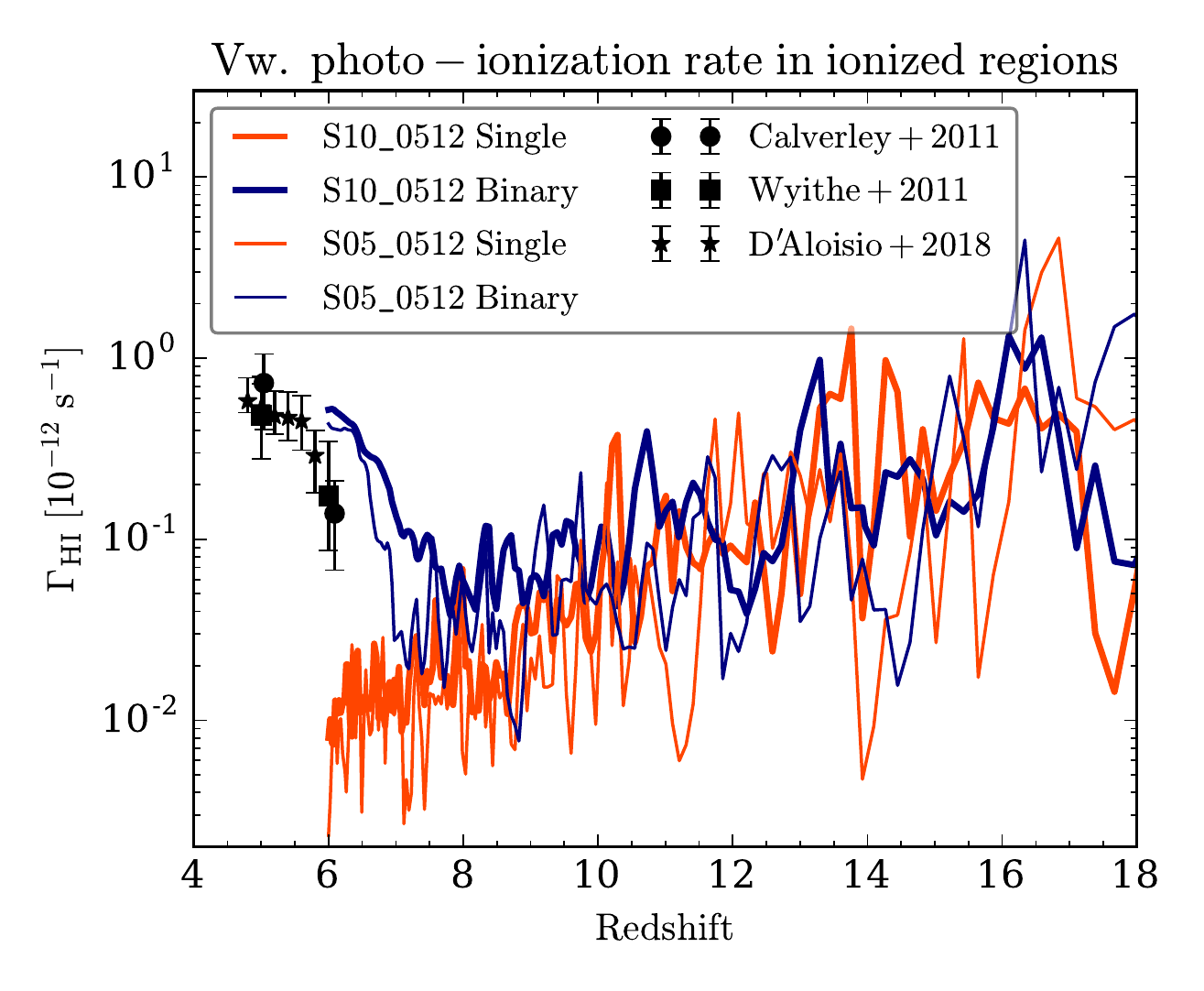}
  \caption
  {\label{Gamma_HI.fig}Redshift evolution of volume-weighted
    photo-ionization rate inside ionized regions ($\xhii>0.5$) in our
    volumes. For comparison we show observational estimates from
    \citet{Calverley2010}, \citet{Wyithe2010}, and \citet{DAloisio2018},
    as indicated in the legend. The rate fluctuates strongly during
    the process of reionization but these fluctuations diminish as the
    volume becomes fully ionized in the binary star simulations. This
    model overshoots observations slightly due to early reionization.}
\end{figure}

In \Fig{Gamma_HI.fig}, we plot the simulated redshift evolution of the
volume averaged $\Gammahi$ in ionized gas, selecting only cells with
hydrogen ionized fraction $\xhii>0.5$. The photo-ionization rate
fluctuates strongly at high redshift, due to the rates being extracted
from a small number of ionized regions nearby or inside haloes.  Thus
$\Gammahi$ is very sensitive the current SFR. Some such regions,
dim yet ionized by previous star formation events, can easily be
identified by matching the temperature and photo-ionization rate
projections in \Fig{map_reionization.fig}. Over time, the number of
ionized regions increases and the strength of the fluctuations
decreases as the average becomes less sensitive to individual
regions. Eventually the fluctuations disappear completely with binary
stars as individual \hii{} bubbles merge and the radiation field in
any location becomes a composite of all the sources in the volume. The
photo-ionization rate slightly over-predicts observations at $z=6$
although this is expected since the box reionizes at a marginally
higher redshift than observations suggest.  For the single stars SED,
the fluctuations remain until $z=6$, since the volume is not fully
reionized, and the average photo-ionization rate is clearly well below
the observational estimates, as expected from the reionization history
in \Fig{xhi.fig}.

We finally note that our use of the variable reduced speed of light
leads to a few tens of percent (relative) over-predictions of the
neutral fraction and under-predictions of the photo-ionization rate,
compared to a full light speed, when the volume is fully ionized (see
\Sec{radidation_setup.sec} and a full explanation in
\cite{Ocvirk2018}: when reionization is completed and the whole volume
becomes optically thin, the photon density turns light-speed
independent and hence both the photon flux and the photo-ionization
rate become light-speed dependent). With the binary SED model, a full
speed of light would thus bring us further away from the $z\approx6$
observational data in Figures \ref{xhi.fig} and \ref{Gamma_HI.fig}.

\subsection{Escape fractions for individual haloes}\label{fesc.sec}

The variable speed-of-light approximation makes it difficult to
estimate $\fesc$ directly from the M1 radiation field, due to the
different and untraceable delay times for photons to travel from their
sources to a given distance. Therefore, we measure the instantaneous
escape fractions in post-processing by tracing rays from all stellar
sources. We calculate the optical depth to neutral hydrogen and
helium, $\tau$, along each ray until it exits the virial boundary of
its parent halo. We then determine the escape fraction for the ray as
$e^{-\tau}$. The net escape fraction from each stellar particle is
then the average of $500$ rays with random directions, and the net
escape fraction for a halo at a given time is the luminosity-weighted
average for all stellar particles assigned to the halo. We assign each
stellar particle to the closest (sub-)halo, using for each halo the
weighted distance measurement $d=r/\Rvir$, where $r$ is the distance
of the star from the halo center. A star outside $\Rvir$ of any halo
is not assigned and we do not assign to any sub-halo fully enclosed
within $\Rvir$ of its parent halo (i.e. stars within such a sub-halo
are assigned to its parent halo).

We have checked in a few
outputs that casting $2000$ rays per source gives relative differences
in $\fesc$ of less than $0.1$ percent, compared to the fiducial $500$
rays. Hence it is fully justified to use $500$ rays per
source. \cite{Trebitsch2017} showed that this method of tracing rays
in postprocessing yields almost identical results as integrating the
flux of M1 photons across the virial radius (with a constant
light-speed) and comparing to the galaxy's ionizing photon production
rate. Note that all of our escape fractions are luminosity-weighted,
meaning $\fesc$ represents the \emph{average escape probability per
  ionizing photon}, not per source.

\begin{figure}
  \centering
  \includegraphics[width=0.5\textwidth]
    {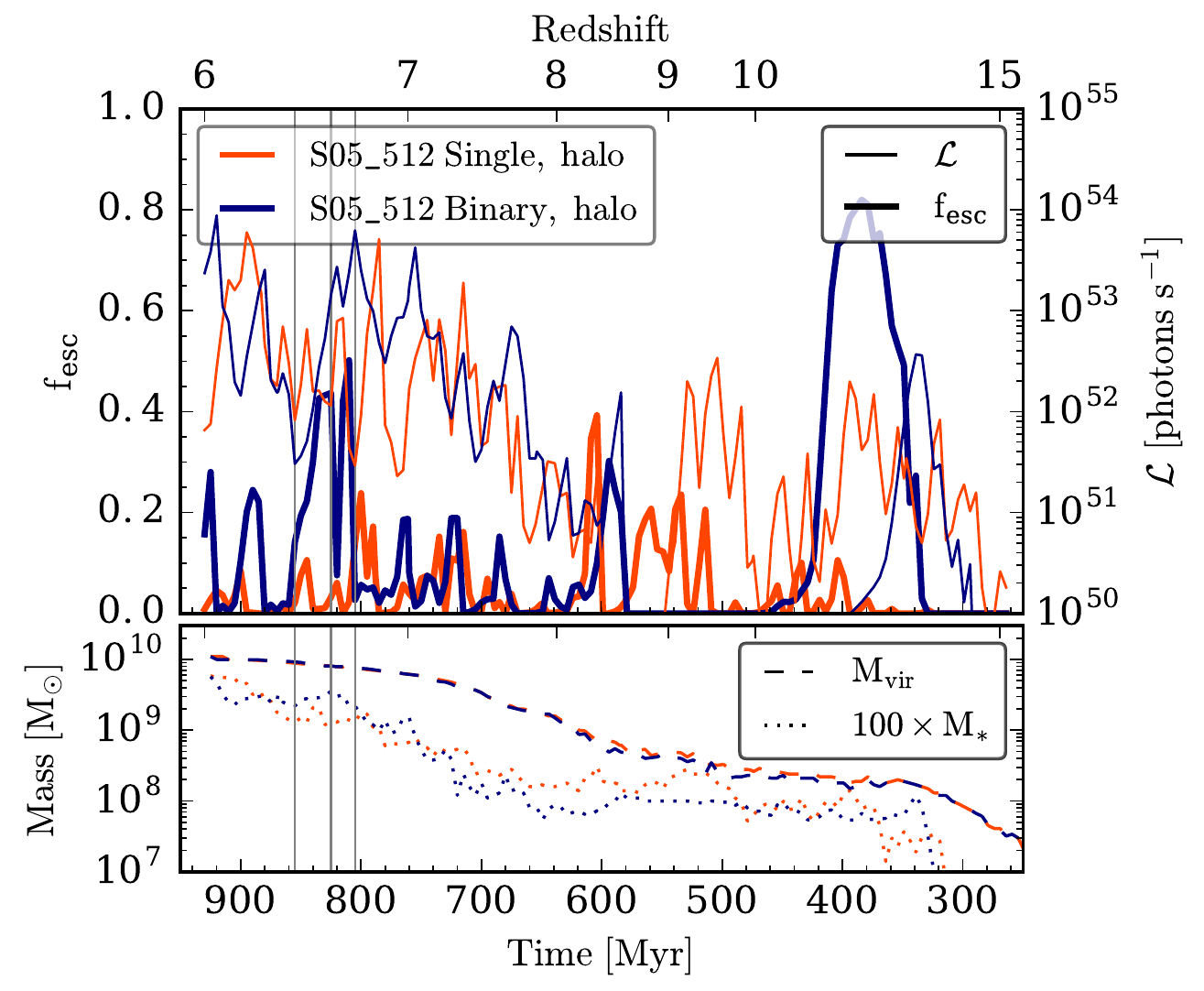}
  \caption
  {\label{fesc_halo.fig}Variations over time in the escape of ionizing
    radiation, $\fesc{}$ (thick solid curves), and intrinsic ionizing
    luminosity, $\IonLum$ (thin solid curves), for the most massive
    halo in the high-resolution $5$ cMpc volume simulations. The red
    curves are with the single-stars SED and the blue curves are with
    binary stars included. In the lower panel, we show the
    corresponding evolution in the virial halo mass (dashed curves)
    and (one-hundred times the) stellar mass, identified as in
    \Sec{galprops.sec}. (Decreases in stellar mass are in part due to
    mass ejected in SN feedback, but more predominantly reflect the
    challenging identification of individual galaxies and their
    assignment to halos, which is especially difficult in the
    merger-rich high-redshift Universe.) Thin vertical solid lines
    mark times corresponding to the halo images in
    \Fig{fesc_halo_maps.fig}. Both $\fesc$ and the production of
    ionizing radiation vary enormously over time, due to feedback
    regulation.}
\end{figure}

\begin{figure*}
  \centering
  \includegraphics[width=0.95\textwidth]
    {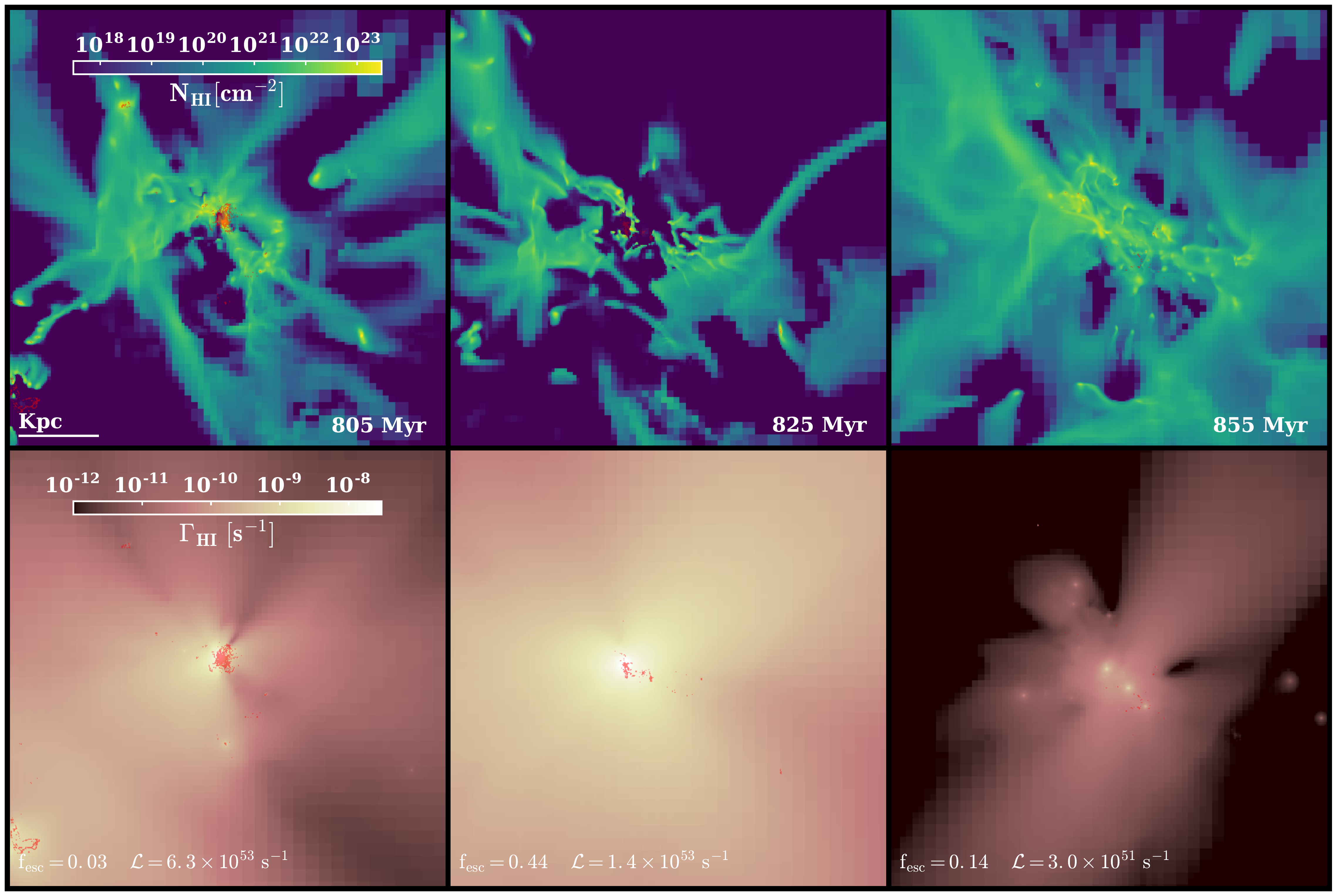}
  \caption
  {\label{fesc_halo_maps.fig}Time series of the main progenitor of the
    most massive halo in the $5$ cMpc volume (binary SED) at
    $z\approx6.8-6.4$ ($\Mvir\approx10^{10} \ \Msun$, $\Rvir\approx9$
    kpc). The upper row shows neutral hydrogen column density and the
    lower row shows the line-of-sight average photo-ionization
    rate. The lower limit for the column density maps, of
    $\Nhi=3 \times 10^{17} \ \ccitwod$, roughly corresponds to a unity
    optical depth for the ionizing radiation. In all panels, stellar
    populations younger than $20$ Myr are superimposed as red
    dots. The three columns of panels correspond to the times
    indicated with vertical lines in \Fig{fesc_halo.fig},
    demonstrating the phases before, during, and after a flash of
    escaping ionizing radiation. The current $\fesc$ and production
    rate of ionizing photons for the halo is written in each
    $\Gammahi$ panel. Initially there is a starburst with a very high
    intrinsic ionizing luminosity, but the ionizing radiation is
    absorbed locally by the dense ISM and only a small fraction of it
    escapes (left). Subsequently, SN feedback disrupts the neutral ISM
    enough so that the radiation efficiently breaks out
    (center). Finally, star formation shuts down as the dense pockets
    of neutral ISM are destroyed. With little SN activity, the
      ISM resettles and absorbs the little ionizing radiation emitted
      from the aging stellar populations with increasing efficiency
      (right).}
\end{figure*}

In \Fig{fesc_halo.fig} we compare the variation over time of $\fesc{}$
from the most massive progenitor to the most massive halo at $z=6$
($\Mvir\approx10^{10} \ \Msun$, $\Mstar\approx10^8 \ \Msun$) in the
smaller-volume \Sphsmhr{} simulations, with single and binary SED
models (in red and blue, respectively).  As found by \cite{Kimm2014,
  Wise2014, Ma2016, Trebitsch2017}, the escape fraction varies
strongly with time, due to regulation by SN feedback. Peaks in $\fesc$
typically follow peaks in ionizing luminosity (denoted by thin
curves) from bursty star formation events. The highly fluctuating
nature of $\fesc$ makes it difficult to compare the binary and stellar
SED models for a single halo, but the plot shows generally higher
$\fesc$ for the binary SED, especially at late times.

To demonstrate how the variation is regulated by feedback,
\Fig{fesc_halo_maps.fig} shows neutral hydrogen column density and
photo-ionization rate projections for the binary SED model at the
three times, indicated by thin vertical lines at $805-855$ Myr in
\Fig{fesc_halo.fig} (at this time, the virial mass of the halo is
$10^{10} \ \Msun{}$). The left column corresponds to a peak of
intrinsic ionizing luminosity, due to an ongoing starburst. The ISM of
the galaxy is already quite disrupted but still intact enough that
little of the ionizing radiation produced can escape. In the middle
column, SN explosions have disrupted the ISM to the extent that star
formation, and hence also the ionizing luminosity, is declining fast,
but the more diffuse ISM now allows a significant fraction
($\fesc=44\%$) of the ionizing photons to break out of the halo. In
the right column, not much ionizing radiation comes from the galaxy
because i) the ISM has settled back and $\fesc$ is very low again and
ii) the galaxy has become very faint in ionizing radiation due to the
drop in star formation. This is a rather extreme example compared to
most other peaks of $\fesc$ in \Fig{fesc_halo.fig}, but the picture is
usually the same, with flashes of escaping ionizing radiation
following star formation (and subsequent feedback) events. The first
such event for the binary case in \Fig{fesc_halo.fig}, at time
$\approx 400$ Myr, is also the most extreme one in terms of duration
and the peak value in $\fesc$. This is because, at this point, the
system is very low in mass ($\Mvir\approx 10^8 \ \Msun$) and highly
susceptible to disruption by SN feedback. This first strong starburst
results in total obliteration of the ISM, so it takes more than a
hundred million years to recover and start forming stars again.

\subsection{Global escape fractions}

\begin{figure*}
  \centering
  \includegraphics[width=0.9\textwidth]
    {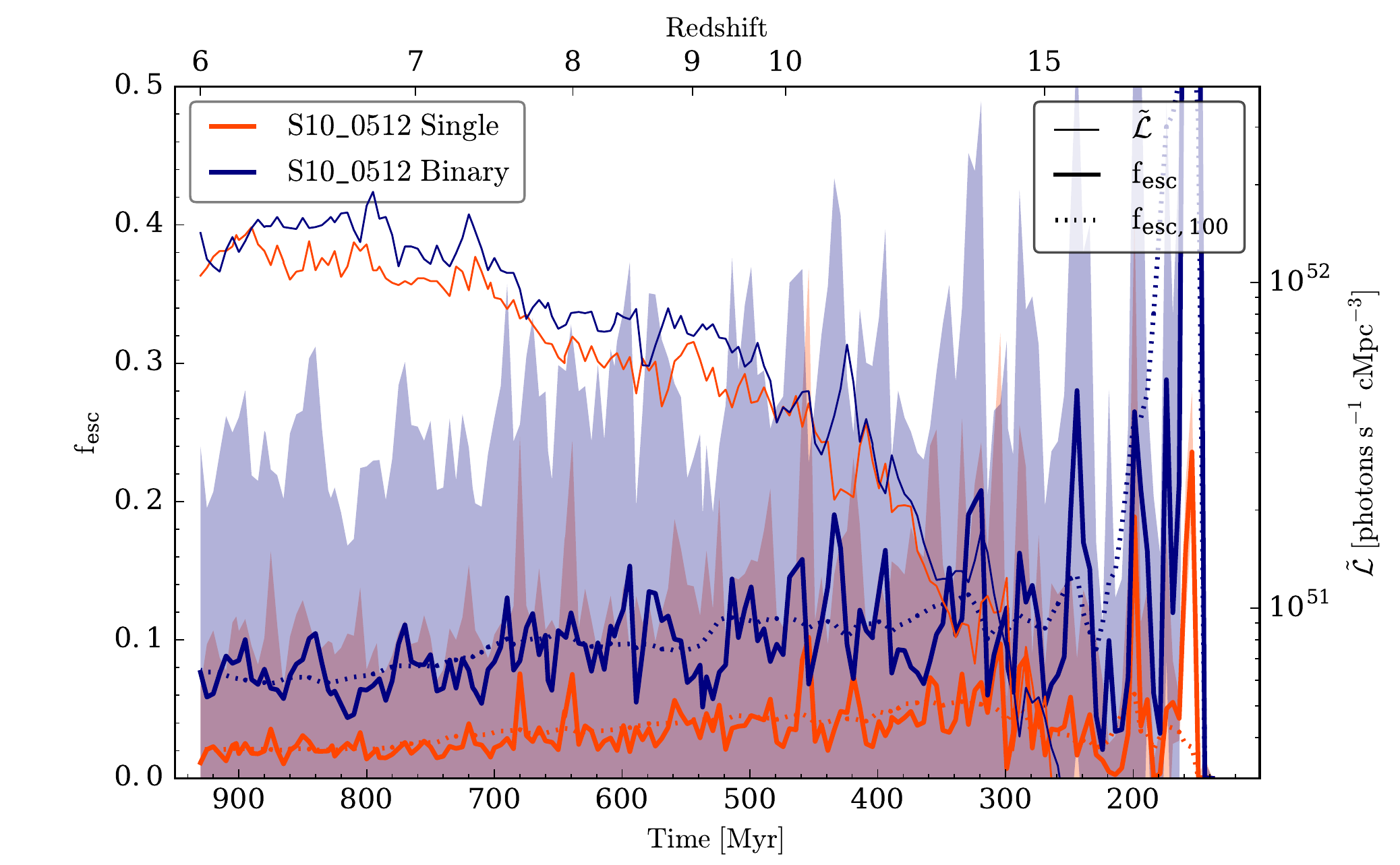}
  \caption
  {\label{fesc_global.fig}Time-evolution of the global
    luminosity-weighted escape fraction $\fesc$ (thick curves; shaded
    regions show the luminosity-weighted variance) and production
    rates per volume for ionizing photons ($\IonLumVol$; thin curves)
    for the $10$ cMpc volume run with single (red) and binary (blue)
    SED models. Dotted lines indicate the luminosity-weighted mean
    $\fesc$ over the last $100$ Myr ($\feschundred$). The binary SED
    model results in systematically higher $\fesc$ compared to the
    single stars SED, by a factor $\approx3$, but also to somewhat
    higher production rates for ionizing photons.}
\end{figure*}

\Fig{fesc_global.fig} shows the global luminosity-weighted $\fesc$ for
the $10$ cMpc volume, again for both the single and binary SED models,
as well as the ionizing luminosities per volume ($\IonLumVol$). We use
the same definition for escape as in the previous section. We remind
that we do not assign particles outside the $\Rvir$ of any halo, but
note that such unassigned stellar particles account for less than
$0.01$ percent of the total ionizing luminosity and hence their
inclusion would have no effect on our estimated escape fractions.

Since the escape fraction is now averaged over a few thousand
galactic sources, the fluctuations are not as extreme as seen in
individual haloes, except at high redshift, when $\fesc$ is dominated
by a small number of galaxies. This makes it easier to compare the
single- and binary-star SED models. In order to make the comparison
even clearer, we plot in dotted curves the $\IonLumVol$-weighted
average over the last $100$ Myr, or $\feschundred$. We show also the
luminosity-weighted variance in the instantaneous $\fesc$, which is
indicated with shaded regions. Note the distribution is highly
non-gaussian and the variance here is meant only to give an indication
of the range in $\fesc$ that we measure at a given redshift.

For the binary SED model, the global escape fraction is at
$\fesc\approx7-10\%$ for $z<9$, while for the single-star SED model it
is at $\fesc\approx2-4\%$. In both cases there is a slight overall
decline in $\fesc$ with decreasing redshift. Sampled over every $5$
Myr until $z=6$, the $\IonLumVol$-weighted average $\fesc$ is $8.5\%$
for binaries and $2.7\%$ for singles (i.e. $\approx3.1$ times higher
for binaries).  This is in good agreement with \cite{Ma2016}, who
predict a factor of $3-6$ boost in `true' escape fractions
(i.e. omitting an extra luminosity boost) with binaries.

The higher $\fesc$ with binaries partially explains the much more
efficient reionization with the binary SED model. However, as shown
with thin curves in \Fig{fesc_global.fig}, the production rate per
volume of ionizing photons, $\IonLumVol$, is also somewhat higher for
the binary SED model. It is the combination of higher escape fractions
and higher ionizing luminosities that leads to earlier reionization
with binary stars. However, since the ratio of escape fractions
between the single and binary SED models (a factor few) is
consistently and significantly higher than the ratio of luminosities
for those same models (typically a few tens of percent), we can
conclude, as did \cite{Ma2016}, that the higher $\fesc$ dominates over
the higher integrated luminosity with binary stars (seen in the lower
panel of \Fig{SEDlums.fig}).
 
\begin{figure}
  \centering
  \includegraphics[width=0.48\textwidth]
    {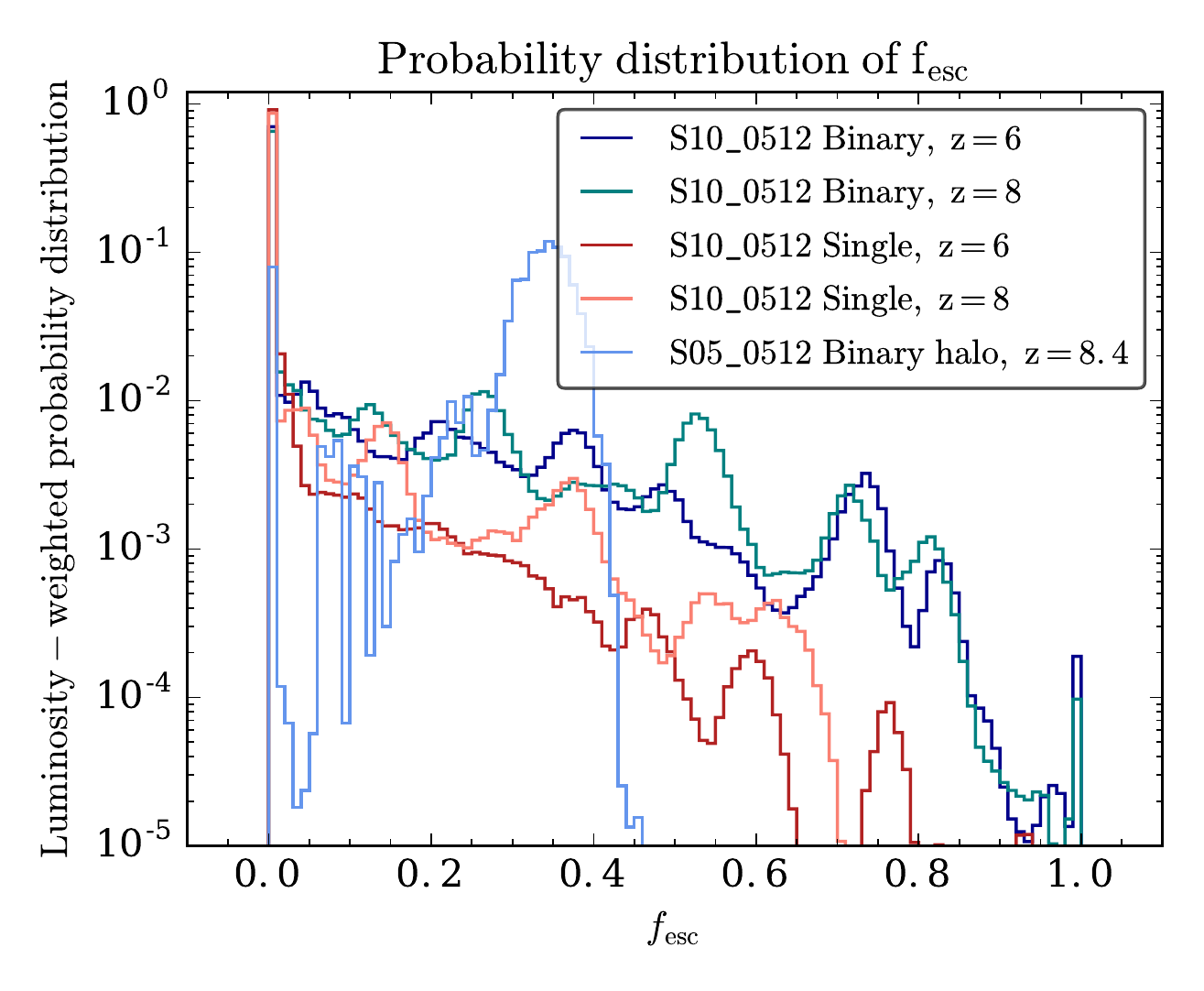}
  \caption
  {\label{fesc_hist.fig} Luminosity-weighted probability distribution
    of escape fractions, $\fesc$, for selected redshifts, comparing
    single and binary SED models, as indicated in the legend. The
    distribution is defined by a `multi-modality', which results from
    a handful of galaxies existing at a given time in a bi-modal
    luminous and optically thin phase, with a similar fraction of the
    sky transparent to the ionizing radiation of each young star. We
    also show the probability distribution for the massive halo with
    the binary SED from \Fig{fesc_halo.fig} at $z=8.4$ ($\approx 600 $
    Myr), when there is a strong peak in $\fesc$: this is to show that
    indeed a single halo with a high $\fesc$ shows a strong bi-modal
    distribution in $\fesc$.}
\end{figure}

We further examine the variation in $\fesc$ in \Fig{fesc_hist.fig},
where we plot the luminosity-weighted probability distribution of
escape fractions from all (halo-assigned) stellar particles at
selected redshifts in our $10$ cMpc wide volume, with binary and
single stars. The distribution peaks at zero or near-zero $\fesc$, as
expected, but for higher $\fesc$ there are a few non-systematic bumps
in each snapshot. The reason for these bumps is that a few galaxies in
each snapshot exist in the luminous and optically thin phase (similar
to the middle column in \Fig{fesc_halo_maps.fig}). In this phase, the
majority of young stars are exposed to the same low column density
channels where a fixed fraction of the ionizing radiation can
escape. In contrast, other stars are embedded in high density regions
with negligible escape fractions \citep[see
also][]{Cen2015,Trebitsch2017}. Such a galaxy typically has a bi-modal
distribution in $\fesc$: we include an example of this in the
light-blue curve in \Fig{fesc_hist.fig}, which shows the $\fesc$
distribution from the massive halo in the \Sphsmhrbp{} simulation at a
time of high $\fesc$ (corresponding to the blue peak at $\approx595$
Myr in \Fig{fesc_halo.fig}). The series of bumps in
\Fig{fesc_hist.fig} represent the rare luminous galaxies with high
$\fesc$, each having a bi-modal $\fesc$ distribution, while most
galaxies at a given time either have a low $\fesc$ and contribute
mostly to the peak at $\fesc\approx0$, or they have a low luminosity
and hence contribute little to the luminosity-weighted distributions
plotted in \Fig{fesc_hist.fig}. With larger simulation volumes, these
bumps would presumably merge into a smooth distribution with a single
peak at $\fesc=0$.

\section{Discussion} \label{Discussion.sec}

With the \sphinx{} simulations, we find a significant boost in escape
fractions with a SED model that includes binary stars, increasing to
$\approx 7-10$ percent escape probability per photon from
$\approx 2-4$ percent without the binaries. With the binary SED model,
the volumes are also fully reionized at $z>6$, even somewhat
prematurely compared to observational constraints. 

This puts our results in mild tension with estimations that a global
$\fesc\gtrsim20\%$ is required to match observational constraints on
reionization \citep[e.g.][]{Kuhlen2012b, Ouchi2017}.  These estimates,
however are based on simple analytic models for the competition
between photo-ionization and recombination that depend on the clumping
factor of gas in the IGM and the intrinsic ionizing luminosity density
in the early Universe, in addition to $\fesc$.  Neither of these two
additional parameters are currently well constrained\footnote{Also,
  the Thompson optical depth \cite{Kuhlen2012b} calibrated their
  results against was a bit higher ($\approx 0.09$) than the current
  estimate ($\approx 0.06$) by Planck.}, so our $\fesc\lesssim20\%$ is
no cause for concern.

However, our results show a degeneracy between the efficiency of
feedback and the reionization history, which is a natural outcome of
the regulation $\fesc$ by feedback, and does appear to generate some
spread in the predicted $\fesc$. In \App{SNcalibration.app} we show
that significantly weaker SN feedback produces very similar
reionization histories as our fiducial strong SN feedback, even if the
rate of star formation, and hence the production of ionizing
photons\footnote{There is not an exact linear relationship between the
  number of stars formed and the number of ionizing photons produced,
  since successive generations of stars have higher metallicities and
  hence lower luminosities (see \Fig{SEDlums.fig}). However, ten times
  more stars certainly produce many times more ionizing photons.}, is
almost an order of magnitude higher.

\begin{figure}
  \centering
  \includegraphics[width=0.5\textwidth]
    {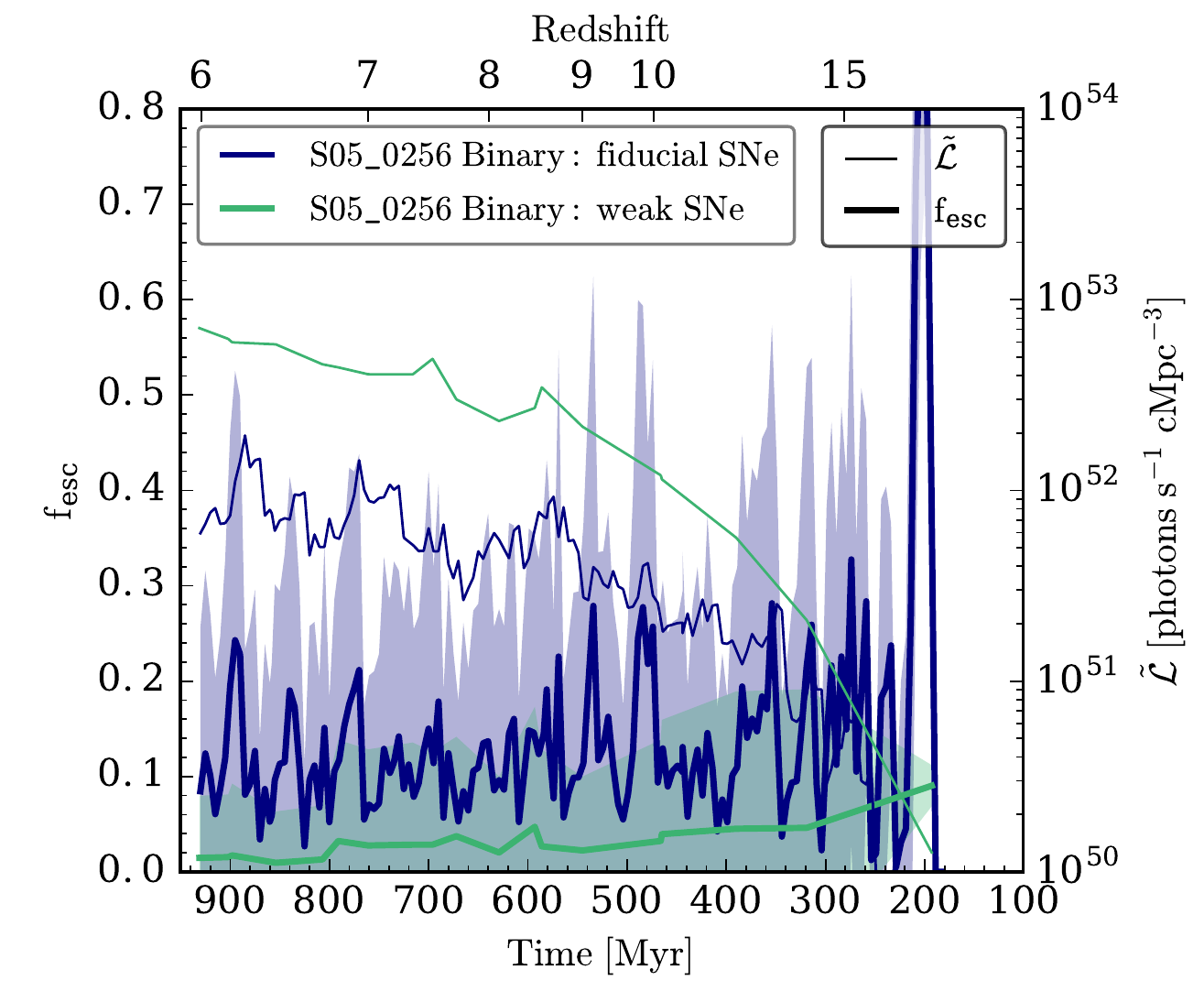}
  \caption
  {\label{fesc_global_SNrate.fig}Time-evolution of the global
    luminosity-weighted escape fractions $\fesc$ (variance in shaded
    regions) and ionizing photon production rates per volume
    $\IonLumVol$ for the $5$ cMpc volume, comparing our fiducial SN
    rate and a weaker SN rate as derived from a Kroupa IMF (in both
    cases using the binary SED model). Note that the frequency of
    \ramses{} outputs is lower in the weak feedback case and hence
    fluctuations in both $\fesc$ and $\IonLumVol$ are likely
    underrepresented. The weaker SN feedback leads to about a factor
    of ten higher production rates of ionizing photons than in the
    fiducial feedback case, but also a factor $\sim 5$ lower
      escape fractions (comparing the luminosity-weighted average for
      $z>6$, which is $\fesc=0.119$ for the fiducial feedback and
      $\fesc=0.023$ with weak SNe).}
\end{figure}

This implies that the global escape fractions are significantly lower
with weak feedback. In \Fig{fesc_global_SNrate.fig} we confirm this
relationship between the strength of feedback and $\fesc{}$. The plot
shows the global luminosity-weighted escape fractions and production
rates of ionizing photons in our $5$ cMpc volume, with our fiducial
rate of SNe and the four times lower SN rate directly derived from a
\cite{Kroupa2001} IMF. Indeed, the typical escape fraction with
fiducial feedback, while highly fluctuating (due to the disrupting
nature of strong feedback and small volume), is much higher than in
the weak feedback case.

In \Sec{galprops.sec} we show that even with our fiducial strong SN
feedback, our galaxy stellar masses (\Fig{smhm.fig}) and luminosity
function (\Fig{flum.fig}) fall towards the upper end of observational
constraints and recent models. The stellar masses can be lowered, via
even stronger feedback, by roughly a factor of two and still be in
good agreement with high-redshift observations and models. Based on
our results with weaker feedback, such enhanced feedback simulations
would likely result in similar reionization histories. Therefore, our
model seems to allow for somewhat higher, but not lower, escape
fractions, while still agreeing with observational constraints of the
high-$z$ Universe.

Our results with binary stars, and in fact the actual event of
reionization itself, appear to contradict observations of the
low-redshift Universe ($z\lesssim1.5$), where indirect measures have
been made for the escape of ionizing photons from dwarf galaxies,
resulting in upper limits of a few percent \citep[e.g.][]{Cowie2009,
  Bridge2010, Siana2010, Leitet2013}. Reconciling this difference
between low observationally estimated escape fractions at low redshift
and relatively high escape fractions required to reionize the Universe
is a well known puzzle which we will not solve in this paper. There
are a few possible explanations that are exciting to explore in future
work. \Fig{SEDlums.fig} demonstrates that the decline in ionizing
luminosity with stellar population age steepens with increasing
metallicity. Similarly, the integrated ionizing luminosities mildly
decrease with increasing metallicity. Both effects suggest a possible
change in SEDs between high and low redshift. If dwarf galaxies become
significantly and increasingly enriched, via long-term local star
formation or external contamination \citep{Angles-Alcazar2017}, their
escape fractions may become very low as a result. Such a scenario,
however, presents some tension with observations, as dwarf galaxies in
the local Universe are not observed to become substantially enriched,
with $\Mstar < 10^8$ $\Msun$ galaxies having $< 1/10$th Solar
metallicity \citep{Kirby2013}.  The evolution in \Fig{fesc_hist.fig}
of our $100$ Myr-averaged escape fraction, $\feschundred$, does hint
at the global escape fraction decreasing slowly over time, but
confirming whether or not this is directly due to
metallicity-dependence in the SED models is beyond the scope of this
paper.

Another factor that may explain the low observed escape fractions is
the stochastic nature of the escape of ionizing radiation from
galaxies \citep[\Fig{fesc_halo.fig},][]{Kimm2014, Ma2016,
  Trebitsch2017} which makes it very difficult to derive an average
escape fraction with only a few observations. Based on a
post-processing of the RHD simulation of \cite{Kimm2014},
\cite{Cen2015} indeed concluded that roughly $\sim100$ galaxy spectra
need to be stacked in order to draw meaningful conclusions on the
average escape fraction.

Tempting as it is, we are reluctant to draw conclusions or preferences
about different SED models from our results. Even if the two SED
models we compare produce very different escape fractions and
interestingly bracket current observational estimates of the
reionization history, neither model produces a very good match with
the observational data in \Fig{Gamma_HI.fig}. While this could be
interpreted as some sort of uncertainty in SED models, it may just as
well be a limitation of unresolved escape fractions. Presumably -- and
this will be attempted in future work -- we could calibrate our
unresolved escape fraction, with $\fescsr\gtrsim1$ for the
single-stars SED model and $\fescsr\lesssim 1$ for the binary stars,
and produce good matches with observations using either SED model.  We
cannot yet say for sure if the escape of radiation from the ISM is
resolved in our simulations, nor can we tell whether it is over- or
under-estimated, and hence it is not timely to rule out any SED
models. We suffice to conclude that in the framework of our models,
the inclusion of binary stars has a huge impact on the reionization
history and makes it easy to reionize the Universe by $z\approx6$.

We do not include active galactic nuclei (AGN) in our simulations, nor
do we consider the effects of molecular cooling or POPIII stars.  AGN
are expected to become relevant in galaxies that are more massive
those currently captured in our simulations \citep[][]{Trebitsch2017b,
  Mitchell2018}. We aim in the next generation of larger-volume
\sphinx{} simulations to study the possible contribution of more
massive galaxies and AGN to reionization. However, their contribution
to the EoR ionizing background is at best uncertain \citep[see the
debate in ][]{DAloisio2016, Chardin2016, Parsa2017}. Regardless, the
inclusion of binary stars in the emission from high-redshift stellar
populations removes the \emph{need} for any other sources of
reionization than stars.

The possible contribution from metal-free (PopIII) stars has been
considered in recent state-of-the-art models \citep[][]{Wise2014,
  Kimm2017} and the general conclusion is that they have relatively
little direct impact on reionization at $z<10$. The impact of PopIII
stars during the EoR is an interesting question, but due to the large
uncertainty in the PopIII initial mass function, a current inclusion
of PopIII physics really constitutes an additional set of free
parameters, which we prefer to skip in the first generation of our
simulation suite.

For similar reasons of simplicity and transparency, we have chosen to
not model the formation and destruction of molecular hydrogen. The
first generation of stars was made possible by primordial molecular
hydrogen formation and cooling, while successive generations formed
via more efficient gas cooling catalysed by metals. In the current
\sphinx{} simulations, instead of modeling the complex molecular
hydrogen formation channels, we have imitated the effects of
primordial molecular hydrogen by calibrating the initial gas
metallicity in our simulations to reach a similar \emph{initial} epoch
and efficiency of star formation as in zoom-simulations where
primordial molecular hydrogen formation was included \citep[using the
models described in ][]{Kimm2017}.

\section{Conclusions} \label{Conclusions.sec}

This paper describes the \sphinx{} suite of cosmological simulations,
the first non-zoom radiation-hydrodynamical simulations of
reionization that capture the large-scale reionization process and
simultaneously start to predict the escape fraction, $\fesc$, of
ionizing radiation from thousands of galaxies. Our series of $5$ and
$10$ co-moving Mpc wide volumes resolve haloes down to the atomic
cooling limit and model the inter-stellar medium of galaxies with
$\sim 10$ pc resolution. We select our cosmological initial conditions
out of $60$ candidates to minimise the impact of cosmic variance on
the volume-emissivity of ionizing photons and to maintain consistent
reionization histories across different volumes.

Owing to strong SN feedback, our haloes are in good agreement with
high-redshift observations, as well as recent state-of-the-art models,
of the stellar mass to halo mass, star formation rate versus halo
mass, and, most importantly, the $1500$ \AA{} UV luminosity function
at $z=6$ (\Fig{flum.fig}).

We study the effect of binary stars on reionization. Due to mass
transfer and mergers between binary companion stars, spectral energy
distribution (SED) models that include binary stars predict higher
integrated ionizing luminosities for stellar populations as well as a
much shallower decline of luminosity with stellar age
\citep[][]{Stanway2016}. We run and compare two sets of \sphinx{}
simulations: one with the \bpass{} SED model \citep{Eldridge2007} that
includes binary stars, and one with the \cite{Bruzual2003} model,
which does not. Using the M1 moment method for radiative transfer, we
inject into our volumes the exact ionizing luminosities dictated by
those SED models, given the metallicity, age, and mass of each stellar
population particle, and without any calibration in unresolved escape
fractions.

The reionization histories strongly differ in the models with and
without binary stars (\Fig{xhi.fig}). The different reionization
histories bracket the observed constraints, with the binaries model
reionizing at $z\approx7$, slightly earlier than observations predict,
while the model with single stars does not ionize the volume by $z=6$.

Because the escape fraction is regulated by feedback, as was
demonstrated by \cite{Kimm2017} and later by \cite{Trebitsch2017}, it
fluctuates strongly with time in an individual galaxy
(\Fig{fesc_halo.fig}), with brief flashes of radiation during which
the galaxy satisfies the two conditions of being luminous and having
an ISM disrupted enough for ionizing radiation to escape. The binary
SED model consistently results in (luminosity-weighted) mean
$\fesc \approx 7-10\%$, with a large scatter. This is a factor
$\approx3$ higher than that for the singles-only SED model
(\Fig{fesc_global.fig}).

The higher escape fractions are due to the combination of feedback and
the shallower decline in stellar luminosities with binaries: by
disrupting the ISM and clearing away dense gas, feedback
simultaneously suppresses star formation and increases the local
$\fesc{}$. With binary stars, more photons can escape, and for a
longer time, after this ISM disruption, leading to larger globally
averaged escape fractions. The higher escape fractions are the primary
reason for the much earlier reionization with binaries, but the higher
integrated luminosity with binaries also plays a sub-dominant role.

Our reionization histories are robust to changes in resolution, volume
size, and, surprisingly, to changes in the SN feedback efficiency. The
robustness to SN feedback is an outcome of an apparent balance between
ionizing luminosities and escape fractions, which scale down and up,
respectively, with increased feedback efficiency
(\Fig{fesc_global_SNrate.fig}).

We will follow up this work with an analysis of how escape fractions
vary with halo mass and how haloes of different masses contribute to
reionization. We will also use the \sphinx{} simulations to study the
back-reaction of suppressed galaxy growth via reionization and to
predict the observational properties of extreme-redshift galaxies,
which will become increasingly visible to us when the JWST comes
online.

\section*{Acknowledgements}
We are grateful the reviewer, John Wise, for comments that
strengthened the paper.  We thank Dominique Aubert, Leindert Boogaard,
Julien Devriendt, Yohan Dubois, Peter Mitchell, Ali Rahmati, Benoit
Semelin, and Maxime Trebitsch for useful discussions, and Andreas
Bleuler for very useful contributions to the optimisation of
\ramsesrt{}.  JR and JB acknowledge support from the ORAGE project
from the Agence Nationale de la Recherche under grant
ANR-14-CE33-0016-03. HK thanks the Beecroft fellowship, the Nicholas
Kurti Junior Fellowship, and Brasenose College. TK is supported by the
National Research Foundation of Korea to the Center for Galaxy
Evolution Research (No. 2017R1A5A1070354) and in part by the Yonsei
University Future-leading Research Initiative of 2017
(RMS2-2017-22-0150). TG is grateful to the LABEX Lyon Institute of
Origins (ANR-10-LABX-0066) of the Universit\'e de Lyon for its
financial support within the program ``Investissements d'Avenir''
(ANR-11-IDEX-0007) of the French government operated by the National
Research Agency (ANR). Support by ERC Advanced Grant 320596 ``The
Emergence of Structure during the Epoch of reionization'' is
gratefully acknowledged. The results of this research have been
achieved using the PRACE Research Infrastructure resource SuperMUC
based in Garching, Germany. We are grateful for the excellent
technical support provided by the SuperMUC staff. Preparations and
tests were also performed at the Common Computing Facility (CCF) of
the LABEX Lyon Institute of Origins (ANR-10-LABX-0066), and on the
GENCI national computing centers at CCRT and CINES (DARI grant number
x2016047376).

\bibliography{library}

\appendix

\section{Luminosity functions} \label{flum.app}

\begin{figure}
  \centering
  \includegraphics[width=0.48\textwidth]
    {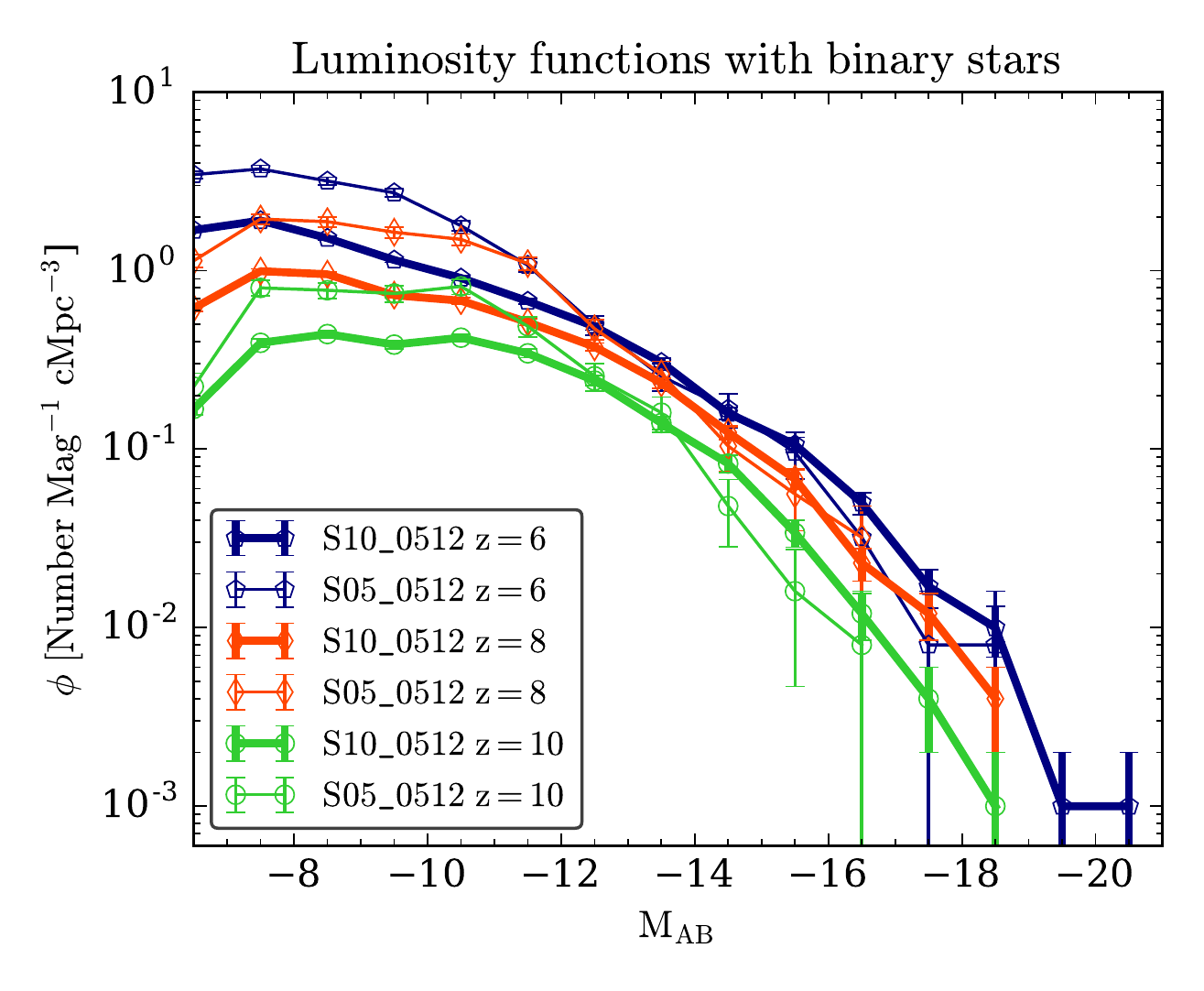}
  \caption
  {\label{flum_allz.fig} Predicted $1500$ \AA{} luminosity functions
    with binary stars at $z=6,8$, and $10$, with Poissonian
    error-bars. }
\end{figure}

For future comparisons to models and/or observations, we show in
\Tab{lumfunc.tbl} our predicted $1500$ \AA{} luminosity functions for
our $5$ and $10$ cMpc wide volumes, with the binary SED, at $z=6,8$,
and $10$. These luminosity functions are also plotted in
\Fig{flum_allz.fig}

\begin{table}
  \begin{center}
  \caption
  {\sphinx{} predictions for the $1500$ \AA{} luminosity function with
    binary stars at $z=6,8$, and $10$. The luminosity functions,
    $\phi_{5}$ and $\phi_{10}$, are from the $5$ and $10$ cMpc wide
    volumes, respectively, and are given in units of
    [$\rm Mag^{-1} \ cMpc^{-3}$].}
  \label{lumfunc.tbl}
  \begin{tabular}{r||ll||ll||ll}
    \toprule
    $\Mab$ & \multicolumn{2}{c||}{$z=10$}
           & \multicolumn{2}{c||}{$z=8$}
           & \multicolumn{2}{c}{$z=6$} \\
           & $\phi_{5}$ & $\phi_{10}$ 
           & $\phi_{5}$ & $\phi_{10}$ 
           & $\phi_{5}$ & $\phi_{10}$ \\
    \midrule
 -20.5   &   0       &	 0 	&   0       &	 0.001 	&   0       &	 0.001 	\\
 -19.5   &   0       &	 0 	&   0       &	 0.001 	&   0       &	 0.001 	\\
 -18.5   &   0       &	 0.001 	&   0       &	 0.004 	&   0.008   &	 0.01 	\\
 -17.5   &   0       &	 0.004 	&   0       &	 0.012 	&   0.008   &	 0.017 	\\
 -16.5   &   0.008   &	 0.012 	&   0.032   &	 0.023 	&   0.032   &	 0.05 	\\
 -15.5   &   0.016   &	 0.034 	&   0.56    &	 0.068 	&   0.096   &	 0.106 	\\
 -14.5   &   0.048   &	 0.083 	&   0.104   &	 0.124 	&   0.168   &	 0.159 	\\
 -13.5   &   0.160   &	 0.140 	&   0.264   &	 0.235 	&   0.256   &	 0.306 	\\
 -12.5   &   0.256   &	 0.242 	&   0.472   &	 0.374 	&   0.496   &	 0.486 	\\
 -11.5   &   0.488   &	 0.344 	&   1.096   &	 0.514 	&   1.072   &	 0.674 	\\
 -10.5   &   0.816   &	 0.421 	&   1.496   &	 0.678 	&   1.784   &	 0.909 	\\
 -9.5    &   0.744   &	 0.385 	&   1.640   &	 0.728 	&   2.728   &	 1.147 	\\
 -8.5    &   0.776   &	 0.441 	&   1.880   &	 0.954 	&   3.176   &	 1.52 	\\
 -7.5    &   0.800   &	 0.394 	&   1.944   &	 0.993 	&   3.712   &	 1.907 	\\
 -6.5    &   0.224   &	 0.168 	&   1.136   &	 0.616 	&   3.448   &	 1.685 	\\
   \bottomrule
  \end{tabular}
  \end{center}
\end{table}

\section{Resolution convergence} \label{res_conv.app}

In this Appendix we compare the main results of our fiducial
simulations to those acquired with lower resolution, in order to show
that they are not significantly affected by minor variations in
spatial and mass resolution.

\begin{figure}
  \centering
  \includegraphics[width=0.48\textwidth]
    {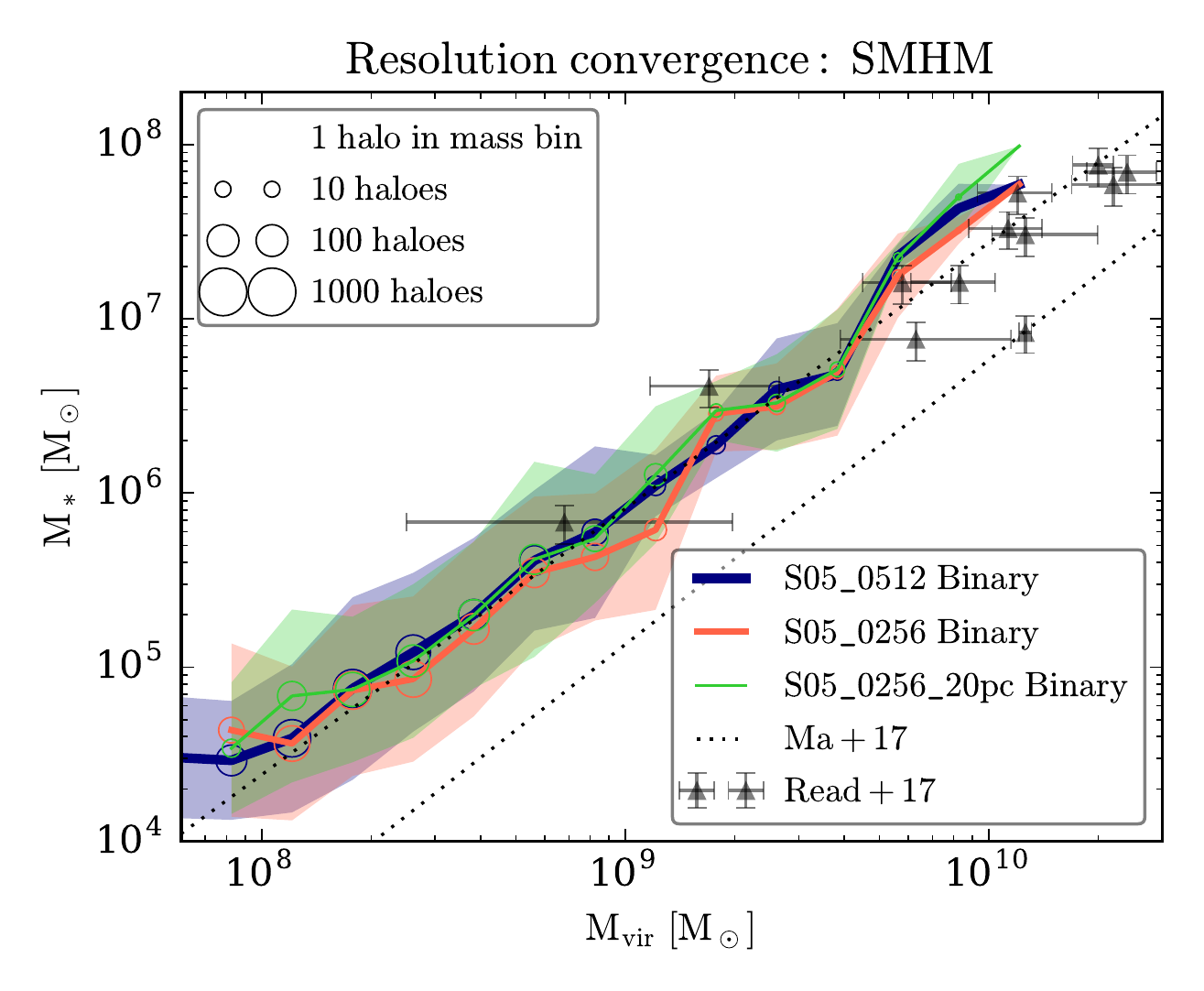}
  \caption
  {\label{smhm_res.fig}Resolution variations of the stellar mass to
    halo mass relation in our $5$ cMpc volume simulations. The
    thickest (blue) curve represents the high-resolution run,
    intermediate thickness (red) represents a run with degraded DM
    mass resolution by a factor of 8, and the thinnest (green) curve
    represents a run with the same degradation in mass and an
    additional degradation of the maximum physical resolution
    ($\approx20$ pc instead of $10$).  Observations and recent
    predictions from simulations are included as in the main
    text. Circles show the number of haloes in each mass bin. The
    changes in resolution have little effect on the SMHM relation.}
\end{figure}

\Fig{smhm_res.fig} shows the $z=6$ stellar mass to halo mass relation
for resolution variations in our $5$ cMpc volume (with the binary SED
model). The thickest (blue) curve shows our high-resolution case,
where the DM mass resolution is $3.1 \times 10^4 \ \Msun$ and the
maximum physical resolution at $z=6$ is $10.9$ pc. The
medium-thickness (red) curve represents a run with our fiducial mass
resolution, where the DM mass resolution is degraded to
$2.5 \times 10^5 \ \Msun$ per DM particle, but the stellar mass and
physical resolution remain fixed. The thinnest (green) curve
represents another run where we have also degraded the maximum
physical resolution, which is now at $21.8$ pc at $z=6$. These
resolution changes have little and non-systematic effects on the
stellar mass to halo mass relation. The most noticeable difference is
a hint of systematically higher stellar mass for the most massive
haloes with degraded physical resolution, but this may as well be due
to low number statistics.

\begin{figure}
  \centering
  \includegraphics[width=0.48\textwidth]
    {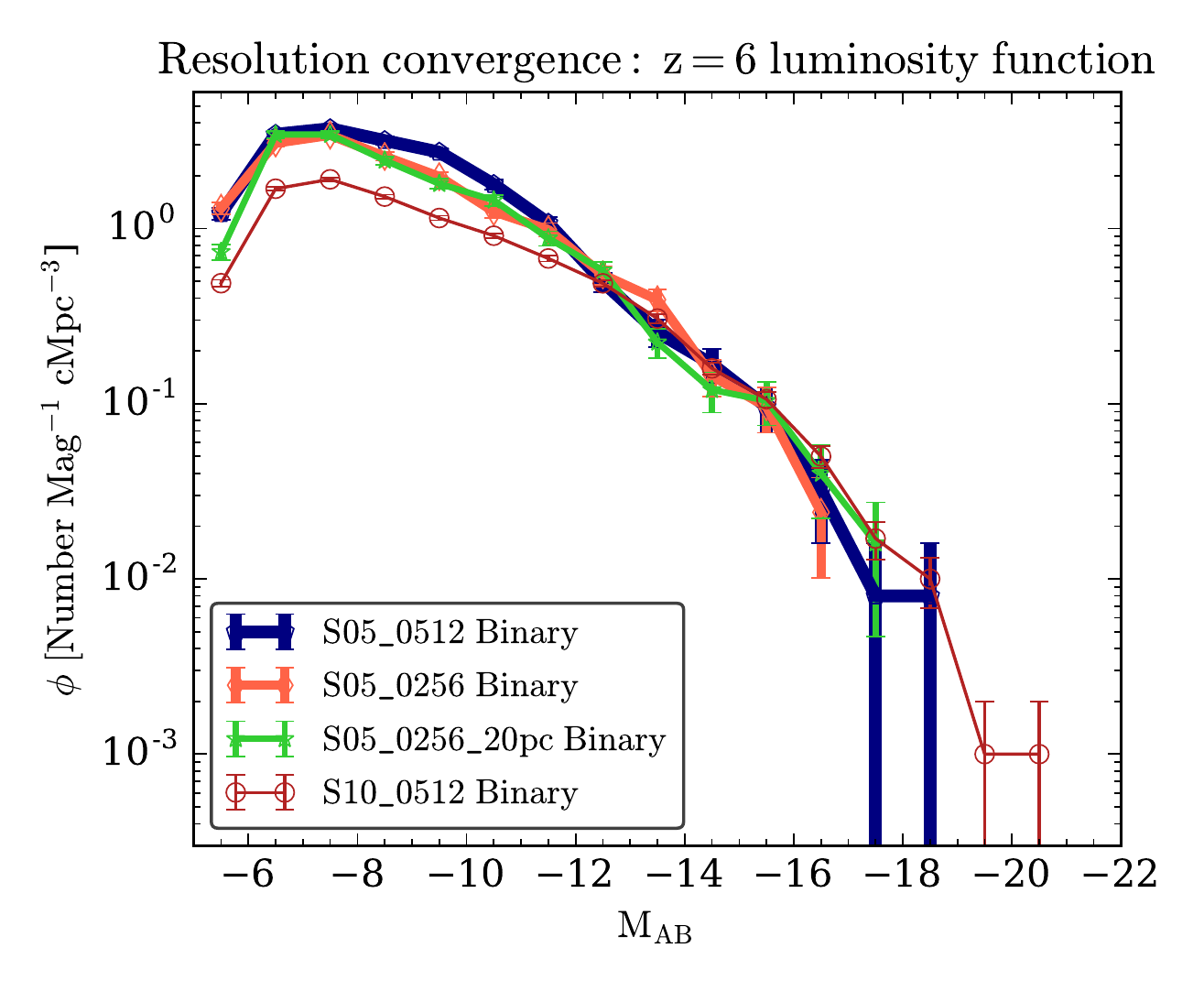}
  \caption
  {\label{flum_res.fig}Resolution variations of the $z=6$ luminosity
    function (with Poissonian error-bars). We show the high-resolution
    run for the $5$ cMpc volume in blue and compare to the same volume
    with degraded mass resolution (orange) and also degraded physical
    resolution (green).  Changing the DM mass resolution and the
    maximum physical resolution has little systematic effect, except
    there is a hint of more galaxies at the faint end with higher DM
    mass resolution. Changing the initial conditions, has a much
    larger effect: in red we show the results for our $10$ cMpc
    volume, with the fiducial resolution matching \Sphsm{}.}
\end{figure}

We show how the luminosity function for the $5$ cMpc volume changes
with resolution in \Fig{flum_res.fig}. There is little systematic
difference for the different resolution runs, except for a hint of
slightly more galaxies at the faint end with higher DM mass
resolution. Comparing with the larger volume (shown in red) the
luminosity function is clearly much more sensitive to the cosmological
initial conditions than to factor-few changes in resolution. From
\Fig{flum.fig} it is also clear that using different SED models has a
larger effect than those resolution changes.  There are, however,
likely stellar mass resolution effects at the low-luminosity end,
manifested in the down-turn in the luminosity functions in the
faintest bin. Our galaxies with $\Mab>-9.5$ contain fewer than $100$
stellar particles on average: their star formation histories are
overly discrete and likely underestimated.

\begin{figure}
  \centering
  \includegraphics[width=0.48\textwidth]
    {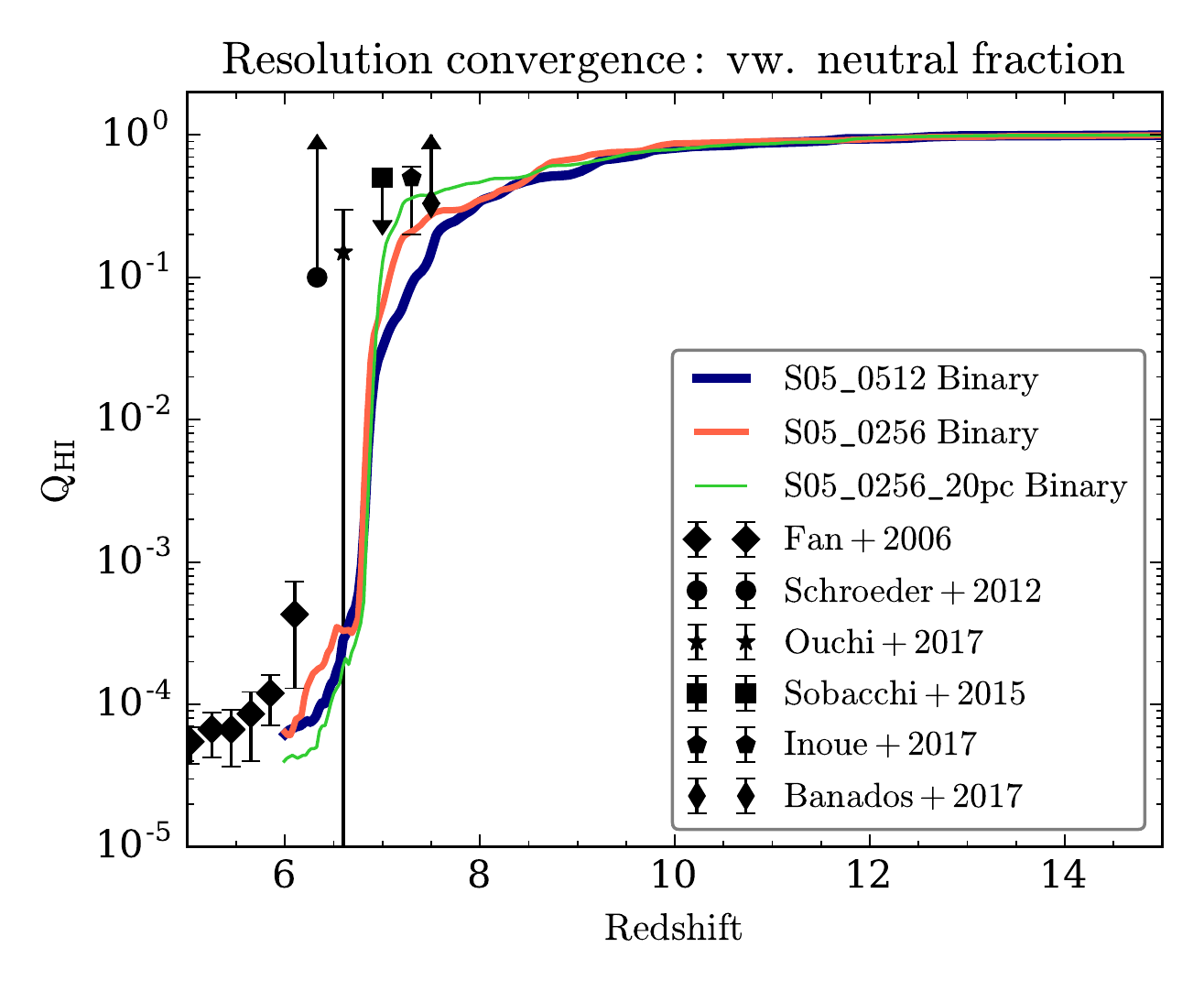}
  \caption
  {\label{xhi_res.fig}Resolution variations of the redshift evolution
    of the volume-filling neutral fraction, $\Qhi$, with the same
    resolution variations (in the $5$ cMpc volume) as in Figures
    \ref{smhm_res.fig} and \ref{flum_res.fig}. Lowering the resolution
    has little effect on the reionization history.}
\end{figure}

In \Fig{xhi_res.fig} we show how the same variations in resolution
affect the reionization history. The aforementioned variations in
resolution have minor effects on the reionization history which do not
appear to be systematic. These minor variations in the $\Qhi$
evolution are likely the result of the stochasticity of starbursts and
feedback in our relatively small simulation volume in the calibration
simulations. We can confirm that simply varying the random seeds for
star formation has similar-amplitude effects on the reionization
history as these resolution variations.

Note that we have not studied a degraded stellar mass resolution. We
expect that variations in the stellar mass resolution within sensible
limits has little effect, since we inject individual SN explosions
rather than instantaneously injecting the total SN energy for the mass
of each stellar population particle (the latter would translate into
stronger resolution effects, with fewer, more energetic SN
explosions). Furthermore, decreasing the stellar particle mass
significantly from our $1000 \ \Msun$ brings us into a questionable
regime where the IMF is no longer sampled by a single particle and we
need to consider more sophisticated models for stellar populations.

\section{Calibration of SN feedback} \label{SNcalibration.app}

We now demonstrate how our SN feedback model is calibrated to match
constraints to the $1500$ \AA{} UV luminosity function and the SMHM
relation.

\begin{figure}
  \centering
  \includegraphics[width=0.48\textwidth]
    {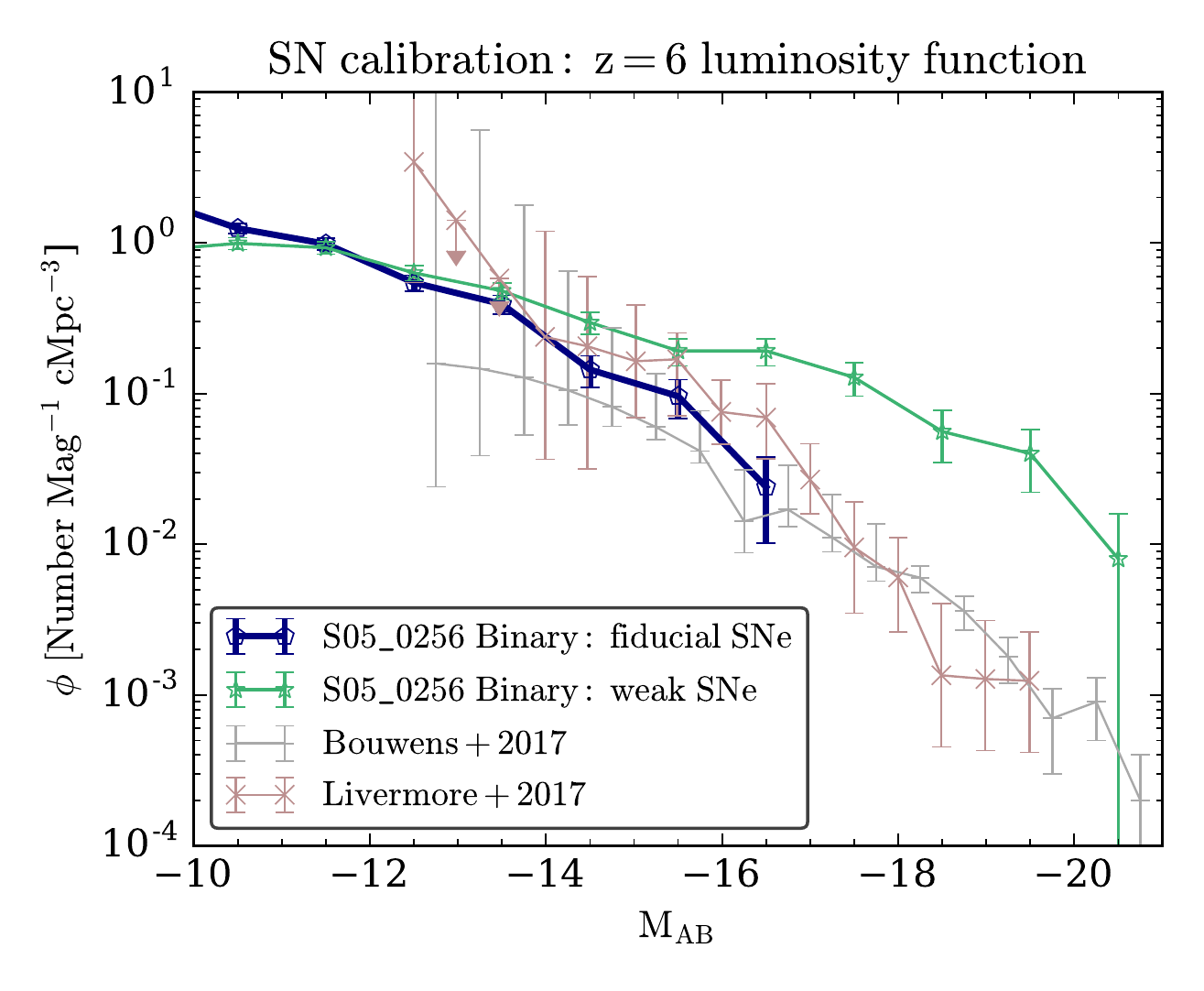}
  \caption
  {\label{flum_SNboost.fig}SN rate calibration effects on the
    predicted luminosity function at $1500$ \AA{}, with two sets of
    $z=6$ observations included for comparison. The error-bars for the
    \sphinx{} data are Poissonian. The `weak' SN rate derived from a
    \citet{Kroupa2001} IMF results in an unrealistic luminosity
    function, while our calibrated, four-times boosted SN rate
    produces good agreement with observations (within the luminosity
    range captured by this $5$ cMpc wide volume).}
\end{figure}

In \Fig{flum_SNboost.fig} we compare simulated $1500$ \AA{} luminosity
functions at $z=6$ in our $5$ cMpc volume (with fiducial resolution)
with observations from \cite{Bouwens2017} and
\cite{Livermore2016}. The thin green curve shows the luminosity
function for an uncalibrated SN rate derived from a \cite{Kroupa2001}
IMF, with $10$ type II SN explosions per $10^3 \ \Msun$. The thick
blue curve shows the results for our fiducial SN feedback, where the
rate of SN explosions has been boosted four-fold, so that $40$ SN
explosions occur per $10^3 \ \Msun$. Both are run with the binary SED
model (which is also used to calculate the luminosity of each galaxy;
as usual dust attenuation is ignored). The lower SN rate is clearly
insufficient to produce a realistic luminosity function, while the
fiducial boosted feedback produces a luminosity function in reasonable
agreement with the observational constraints. As we have seen in
\Fig{flum.fig}, this agreement for the fiducial calibrated SN
feedback, holds both in the case of higher mass resolution and for an
eight times larger volume.

\begin{figure}
  \centering
  \includegraphics[width=0.48\textwidth]
    {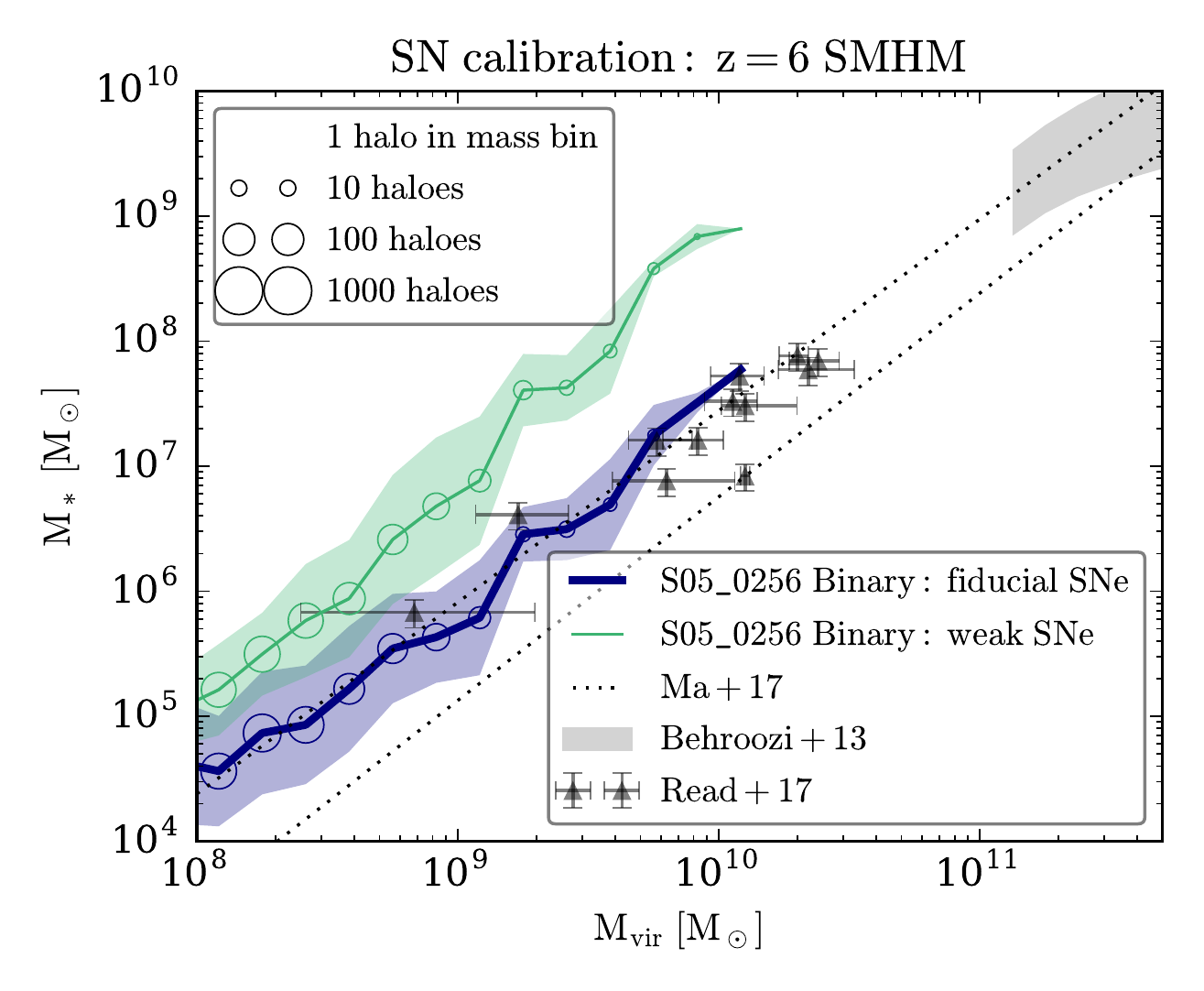}
  \caption
  {\label{smhm_SNboost.fig}Stellar mass to halo mass relation for our
    fiducial and weak, un-calibrated, SN feedback, as indicated in the
    legend. The sizes of the circles indicate the number of haloes in
    each mass bin and the shaded regions indicate the standard
    deviation. Calibrated SN feedback results in a good agreement with
    (local Universe) observations.}
\end{figure}

\Fig{smhm_SNboost.fig} shows a comparison, for the same two
simulations, of the predicted SMHM relation. We include the local
Universe observations from \cite{Read2017} and the $z=6$ relation
derived by \citet{Behroozi2013}. Our fiducial SN feedback results in a
SMHM relation that agrees well with local Universe observations and does
not appear in conflict with high-z constraints. The uncalibrated
feedback produces much higher stellar masses, by almost an order of
magnitude at the high-mass end.

Our fiducial four-fold boost in the rate of SN explosions is a simple
substitute for an unknown combination of the following and possibly
incomplete list of factors: i) uncertainty in the IMF and hence the
rate of SN explosions. ii) resolution, which may play a role, but
likely not a large one, as our resolution has been demonstrated to
produce results that are resolution-converged and not subject to
severe over-cooling. iii) uncertainty in gas cooling rates, which
affects the final momentum gained in the remnants of SN
explosions. iv) lack of complementary feedback physics, such as cosmic
rays, stellar winds, or active galactic nuclei, though the latter is
unlikely in the mass range for galaxies captured in our simulations
\citep{Trebitsch2017b}.

\begin{figure}
  \centering
  \includegraphics[width=0.48\textwidth]
    {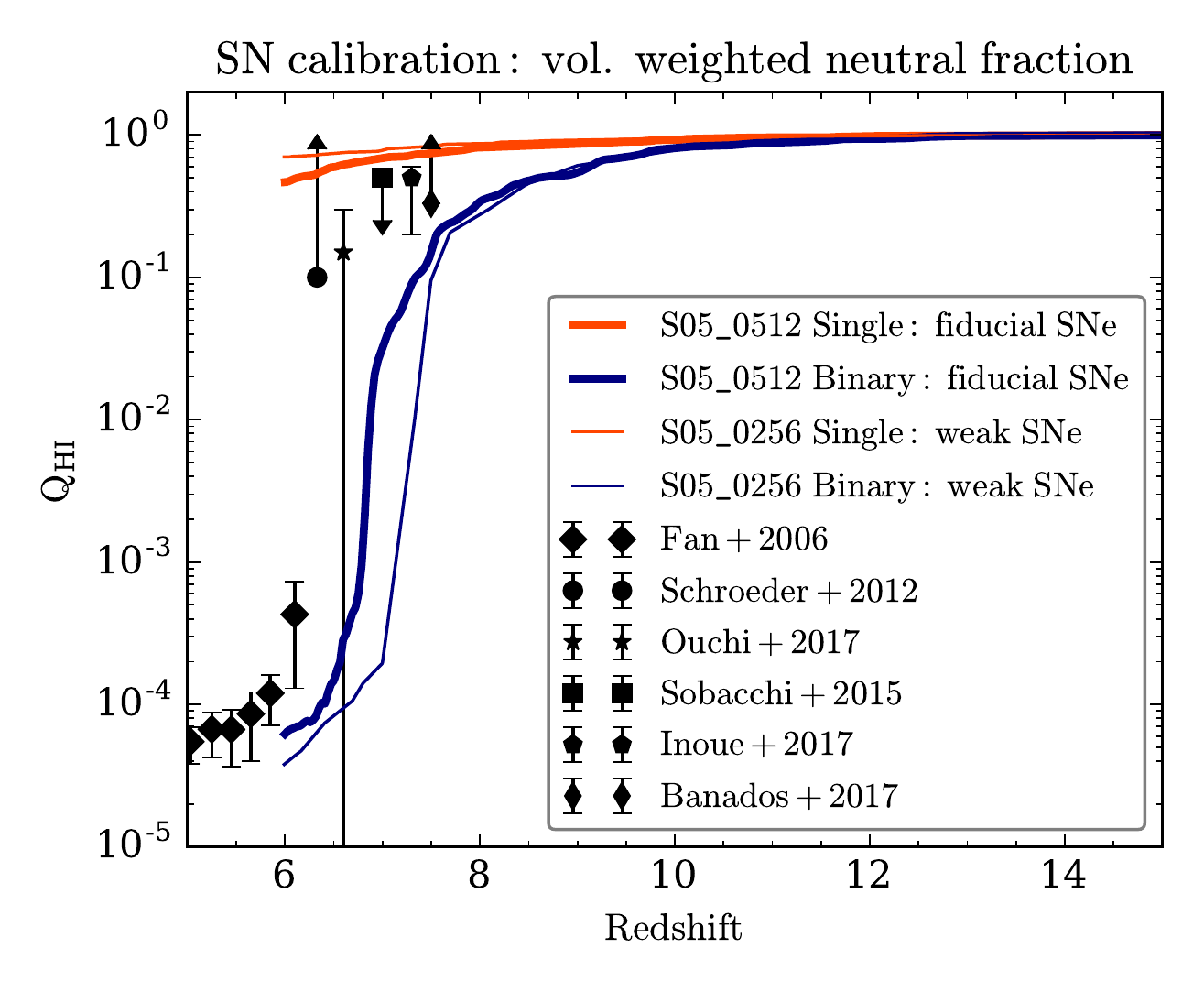}
  \caption
  {\label{xhi_SNres.fig}Calibration effects on the redshift evolution
    of the volume-filling neutral fraction, $\Qhi$. We show the $5$
    cMpc volume run with both single and binary SED models, as
    indicated in the legend, as well as with our fiducial and weak SN
    feedback, and compare to the same observational estimates as in
    \Fig{xhi.fig}. Reducing the rate of SNe by a factor of four has
    little effect, neither on the reionization history nor the very
    significant speed-up resulting from the inclusion of binary
    stars.}
\end{figure}

Regardless, our main result -- of a large difference in the
reionization histories produced by single and binary stellar
populations -- is robust with respect to the feedback
calibration. \Fig{xhi_SNres.fig} shows that the reionization histories
are very similar if the feedback is not boosted, and the same
difference remains between single and binary stellar populations. It
may seem counter-intuitive that reionization progresses as efficiently
with weak SN feedback, but this is due to the feedback regulation of
both star formation and $\fesc$. Many more stars are produced with the
weaker feedback model and this leads to larger ionizing luminosities.
However, the SNe also become less efficient at disrupting the ISM,
leading to a smaller fraction of photons which escape into the
IGM. With the higher luminosities but lower escape fractions, a
comparable number of ionizing photons escapes the galaxies compared to
our fiducial model,

\section{SED properties} \label{SEDprops.app}

The ionizing luminosities of stellar particles in
our simulations, as well as the properties of the three radiation
groups, are based on two spectral energy distribution (SED) models,
which each provide a set of spectra binned by metallicity and
population age (assuming an instantaneous formation time), and
normalised to $\Msun$.

\begin{figure*}
  \centering
  \subfloat
  {\includegraphics[width=0.53\textwidth]
    {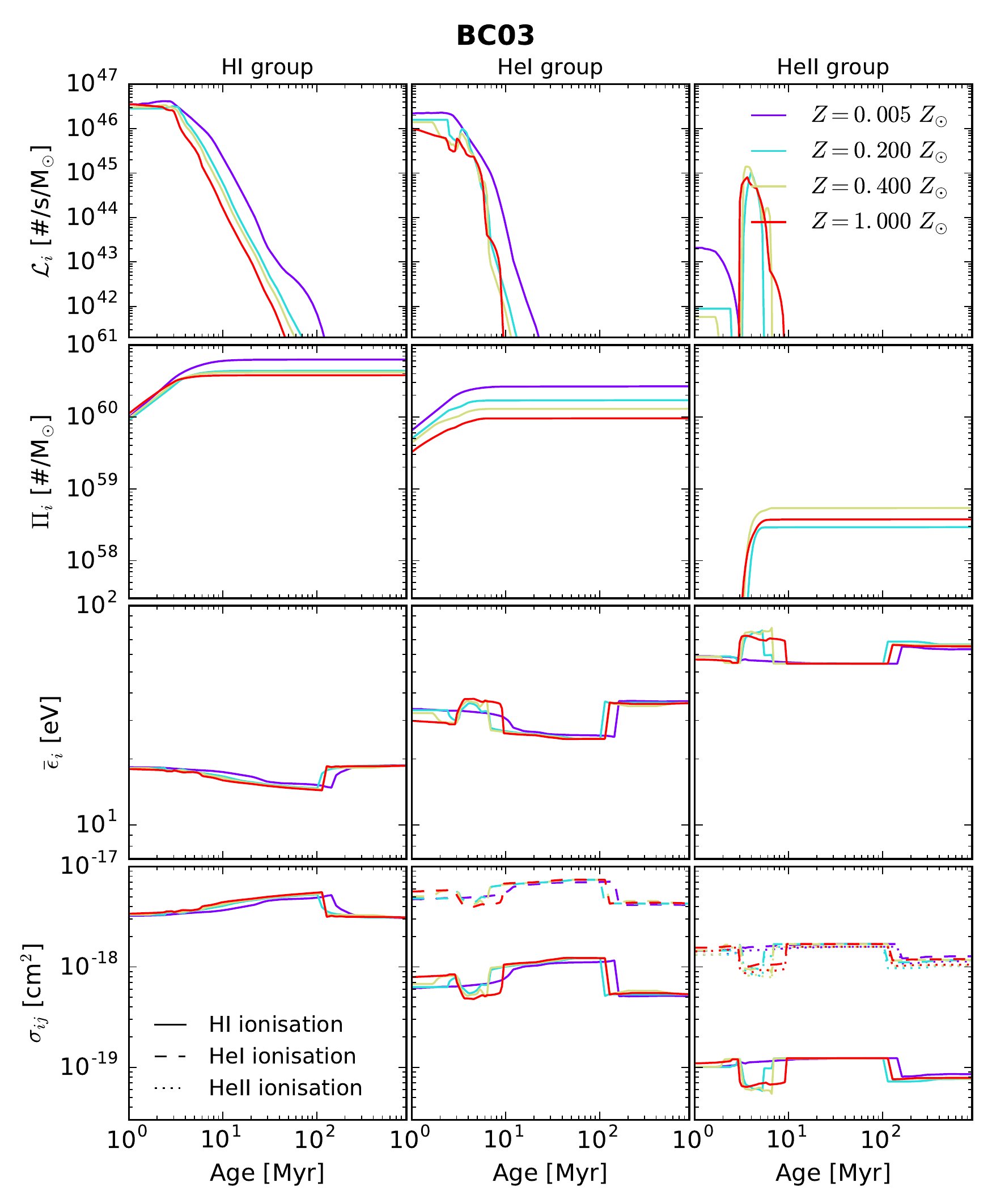}}
  \hspace{-13mm}
  \subfloat
  {\includegraphics[width=0.53\textwidth]
    {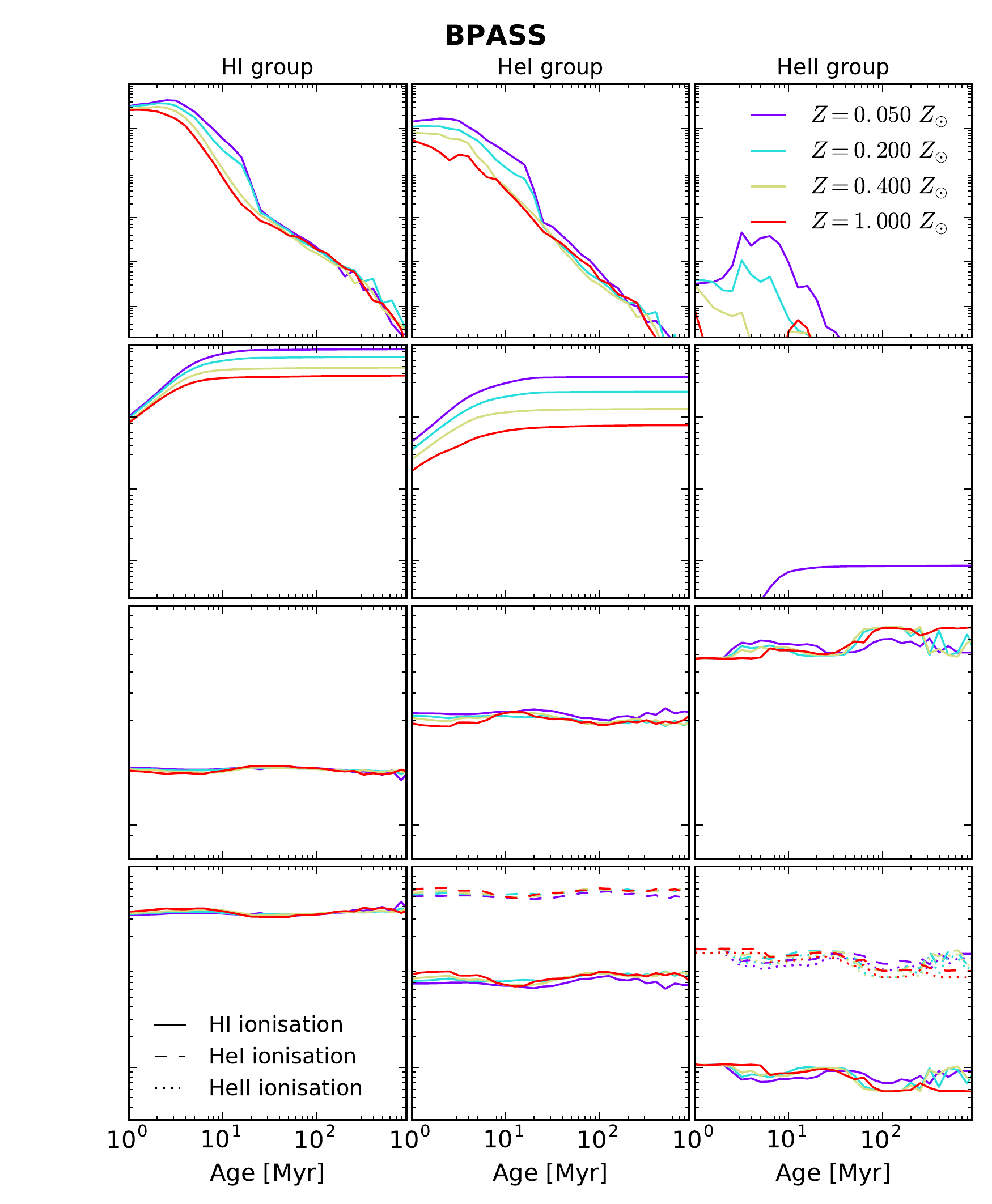}}
  \caption
  {\label{SEDs.fig}Properties of the two spectral energy distribution
    models used in this work. BC03, on the left, is the single star
    model from \citet{Bruzual2003}, while BPASS, on the right, is the
    model from \citet{Eldridge2007}, which includes binary stars. For
    each SED model, the three columns are for the three radiation
    groups (\hi, \hei, and \heii-ionizing), defined in
    \Tab{groups.tbl}. The rows, from top to bottom, show the stellar
    age evolution of the ionizing luminosity $\mathcal{L}_i$ (number
    of ionizing photons in radiation group $i$ per Solar mass per
    time), the integrated luminosity $\Pi_i$ (total number of ionizing
    photons emitted in group $i$ per Solar mass), the average energy
    per photon, $\epsilon_i$, over the energy range defined by group
    $i$, and photon number-weighted average cross sections,
    $\sigma_{ij}$, over the energy range defined by group $i$ for
    photo-ionizing species $j$ (in \hi, \hei, and \heii), using the
    frequency-dependencies for photo-ionization from
    \citealt{Verner1996}. In each plot, the colours denote the
    metallicity assumed for the stellar population, as indicated in
    the legend in the top right panel for each SED model.}
\end{figure*}

One set of simulations uses \cite{Bruzual2003} (\bc{}), which includes
only single star systems. Here, we use the model generated with the
semi-empirical BaSeL 3.1 stellar atmosphere library
\citep{Westera2002} and a \cite{Kroupa2001} initial mass function
(IMF). The other set of simulations uses the \bpass{} model
\citep{Eldridge2007}, which includes binary stars, and has an IMF
closest to \cite{Kroupa2001}, with slopes of $-1.3$ from $0.1$ to
$0.5 \ \Msun{}$ and $-2.35$ from $0.5$ to $100 \
\Msun$. \Fig{SEDs.fig} shows in detail the age- and
metallicity-dependency of the stellar population luminosities, number
weighted mean photon energies, and cross sections, for all three
radiation groups (defined in \Tab{groups.tbl}), integrated from the
two SED spectra for bins of population age and metallicity.

From each SED spectrum (in units of
$\rm{erg \ s^{-1}} \ \Msun^{-1} \ \AA^{-1}$), we extract and tabulate
age- and metallicity-dependent ionising luminosities, average
energies, number-, and energy-weighted cross sections for all
radiation groups, by integrating over the wavelength interval defined
for each group.\footnote{For BC03, the ages range from zero to $13.8$
  Gyr, and for BPASS from zero to 20 Gyr. The metallicities range from
  $0.005-5 \ \Zsun$ for BC03 and $0.05-2 \ \Zsun$ for BPASS}. Using
linear interpolation, we then re-bin the tables in age and metallicity
(8 bins) so that the increment is linear in log-space, i.e.
$\Delta log\left(age/Gyr\right)=0.02$ is the same for every
age-interval (giving 266 bins for BC03, 209 bins for BPASS) and
$\Delta log\left(Z/\Zsun\right)$ ($=0.43$ for BC03, $0.23$ for BPASS)
is the same for every metallicity-interval. The purpose of the
re-binning is to speed up the age- and metallicity-interpolation
performed for every star particle in every RHD time-step.  The lowest
tabulated zero-age is a special case for the interpolation, since we
cannot tabulate the logarithm of zero.

For each re-binned age bin, we integrate and tabulate the cumulative
luminosity (number of photons per Solar mass emitted up to given age),
assuming constant metallicity. We then use this re-binned table to
extract the number of photons emitted by a stellar particle by
comparing the cumulative luminosity before and after the RHD
timestep. We use the cumulative luminosity rather than simply
multiplying the instantaneous luminosity by the timestep length,
because if the timestep happens to be very large, short bursts in
luminosity may be lost entirely. Using the cumulative luminosity
ensures that these unresolved bursts are still accounted for in the
photon budget.

For metallicities outside the table limits, we use the lowest or
highest tabulated metallicity (this is irrelevant for age, which is
always within the tabulated limits). We feel this is a more
conservative option than to apply extrapolation outside the
metallicity limits: the BC03 model reaches ten times lower metallicity
than BPASS, and lower metallicity generally leads to higher ionising
luminosity (see \Fig{SEDs.fig}). Hence, if there is an under-estimate
of luminosities at very low metallicities due to our lack of
extrapolation, the under-estimate should be more exaggerated with
BPASS than with BC03, i.e. we are then underestimating the difference
between BPASS and BC03.

Every ten coarse time-steps, we update the photo-ionization cross
sections and energies for the radiation groups in our simulations, to
represent the luminosity-weighted average of all emitting sources in
our simulations. The details of this process are described in Appendix
B2 in \cite{Rosdahl2013}. Note that since stellar population particles
have different ages and metallicities, and hence different spectra as
defined by the assumed SED model, the emission of photons into a
radiation group cannot be exactly energy-conserving and photon-number
conserving at the same time. Given the energy per photon defined for
the group, we can choose to emit either i) the exact amount of energy
per time, leaving a small error in the number of photons per time, due
to the difference in the particle spectrum over the group frequency
interval from the `average' group spectrum, or ii) the exact number of
photons, sacrificing in the same way the exact energy luminosity. In
this work, we have chosen the latter, as the number of ionizing
photons is more important in the context of reionization.

\end{document}